\documentclass[aps,pre,showpacs,
amssymb,twocolumn
]{revtex4-2}
\usepackage{graphicx}
\usepackage{amsmath}
\usepackage{amsfonts}
\usepackage{amssymb}
\usepackage{epsfig,libertine}
\usepackage[libertine]{newtxmath}
\usepackage{color}
\usepackage{dsfont, hyperref, xcolor}
\usepackage{comment}
\usepackage{enumerate}

\usepackage{amsthm} 

\newcommand{\abs}[1]{\left\vert#1\right\vert}

\newcommand{\Tr}[1]{\text{Tr}\left\{#1\right\}}
\newcommand{\ParTr}[2]{\text{Tr}_{#1}\left\{#2\right\}}
\newcommand{\bra}[1]{\langle#1\vert}
\newcommand{\ket}[1]{\vert#1\rangle}

\newtheorem{theorem}{Theorem}

\begin{document}

\title{Optimal work extraction from quantum batteries based on the expected utility hypothesis}

\author{Gianluca~Francica and Luca Dell'Anna}
\address{Dipartimento di Fisica e Astronomia e Sezione INFN, Università di Padova, via Marzolo 8, 35131 Padova, Italy}

\date{\today}

\begin{abstract}
Work extraction in quantum finite systems is an important issue in quantum thermodynamics. The optimal work extracted is called ergotropy, and it is achieved by maximizing the average work extracted over all the unitary cycles. However, an agent that is non-neutral to risk is affected by fluctuations and should extract work by following the expected utility hypothesis. Thus, we investigate the optimal work extraction performed by a risk non-neutral agent by maximizing the average utility function over all the unitary cycles. We mainly focus on initial states that are incoherent with respect to the energy basis, achieving a probability distribution of work. In this case we show how the optimal work extraction will be performed with an incoherent unitary transformation, namely a permutation of the energy basis, which depends on the risk aversion of the agent. We give several examples, in particular also the work extraction from an ensemble of quantum batteries is examined. Furthermore, we also investigate how work extraction is affected by the presence of initial quantum coherence in the energy basis by considering a quasiprobability distribution of work.
\end{abstract}

\maketitle
\section{Introduction}
Work extraction in isolated finite quantum systems is performed by a cyclical change of the Hamiltonian parameters (a unitary cycle). For instance, these quantum systems can be used to temporarily store energy, and thus they can be used as quantum batteries in a consumption center where an agent will extract useful work from them (see, e.g., Ref.~\cite{campaioli23} for a review). Work is typically affected by stochastic fluctuations so that the extraction is not deterministic.
Thus, typically an agent aims to  extract an amount of work making a choice (among the different unitary cycles) under uncertainty.
When the agent is neutral to risk, he will choose the unitary operator which maximizes the average work extracted.
In this case, the optimal work extracted is given by the ergotropy and the optimal unitary cycle leads the initial state of the system to a passive state~\cite{Allahverdyan04}. However, if the agent takes in account the risk of his choice, he can make his choice relying on the expected utility hypothesis, first formalized by von Neumann and Morgenstern within the theory of games and economic behaviour in 1944~\cite{vonNeumann}.
Depending on the risk aversion of the agent, the latter can choose a unitary cycle that leads to a nonpassive final state. Formally, risk aversion can be characterized in terms of a utility function, which quantifies the satisfaction gained from a choice.
Thus, the work extracted can be thought of as wealth and the utility can be used to select the unitary cycle.
Furthermore, when the work satisfies some fluctuation theorems, these can influence the expected utility and the choice performed by the agent~\cite{Ducuara23,Francica23}.  In particular, in Ref.~\cite{Francica23}, we investigated how an entropy coming from a detailed fluctuation theorem can influence the decision of the agent. Here, we focus on a completely different problem that is the optimal work extraction performed by a risk non-neutral agent, generalizing the concept of ergotropy by looking on the expected utility hypothesis.
We aim to answer to the questions: What is the optimal unitary cycle of work extraction when the expected utility hypothesis is taken in account? How does this differ from ergotropy? How to select between different initial states that provide different work extraction statistics?
To answer to these questions, in Sec.~\ref{sec.preliminaries} we introduce some preliminary notions. In Sec.~\ref{sec.ergotropy} we review the concept of ergotropy, corresponding to the optimal work extraction by an agent neutral to risk. In Sec.~\ref{sec.utility} we introduce some rudiments of expected utility theory. To get a probability distribution of work, we focus on initial states which are incoherent with respect to the energy basis.
Then, we study the optimal work extraction by an agent non-neutral to risk in Sec.~\ref{sec.optimalinco}.
We mainly focus on an exponential utility function, from a general point of view in Sec.~\ref{sec.exputi}, and by considering particular cases which are a qubit in Sec.\ref{sec.exputid2}, a qutrit in Sec.~\ref{sec.exputid3}, two qubits in Sec.~\ref{sec.exputid2+2} and an ensemble of quantum batteries in Sec.~\ref{sec.ensemble}. We also discuss the case of an arbitrary utility function in Sec.~\ref{sec.arbiuti}. In particular, we prove that the optimal expected utility function is achieved by a cycle that is an incoherent unitary transformation in the energy basis. Furthermore, in Sec.~\ref{sec.coe} we briefly investigate how the expected utility (by considering an exponential utility function) is affected by the presence of initial quantum coherence in the energy basis. When the initial state is not incoherent, there may not be a probability distribution for the work done, as proven by a no-go theorem~\cite{Perarnau-Llobet17}. This is related to the quantum contextuality as discussed in Ref.~\cite{Lostaglio18}. Thus, to perform our study we adopt the quasiprobability distribution of work introduced in Ref.~\cite{Francica22}, which is selected if some fundamental conditions need to be satisfied~\cite{Francica222}.

\section{Preliminaries}\label{sec.preliminaries}
We start our discussion by introducing some preliminary notions, which are the ergotropy (see Sec.~\ref{sec.ergotropy}) and some rudiments about the expected utility theory (see Sec.~\ref{sec.utility}).
\subsection{Ergotropy}\label{sec.ergotropy}
Following Ref.~\cite{Allahverdyan04}, we focus on a quantum system with a finite Hilbert space of dimension $d$. The system is prepared in the initial state
\begin{equation}
\rho = \sum_{k=1}^d r_k \ket{r_k}\bra{r_k}
\end{equation}
and with the initial Hamiltonian
\begin{equation}\label{eq.hami}
H = \sum_{k=1}^d \epsilon_k \ket{\epsilon_k}\bra{\epsilon_k}\,,
\end{equation}
where $r_k\geq r_{k+1}$ and $\epsilon_k< \epsilon_{k+1}$. The system is thermally isolated, and an agent extracts an amount of average work by  cyclically changing some Hamiltonian parameters, so that at the end of the cycle the final Hamiltonian is equal to the initial one $H$. Thus, the agent perform a unitary cycle with a unitary time-evolution operator $U=\mathcal T e^{-i \int_0^\tau H(t)dt}$ generated by the time-dependent Hamiltonian $H(t)$ such that $H(0)=H(\tau)=H$, where $t=0$ and $t=\tau$ are the initial and final time, and $\mathcal T$ is the time ordering operator. The final state is $U\rho U^\dagger$ and the average work is minus the change of average energy and reads
\begin{equation}\label{eq.workave}
W(\rho,U)=E(\rho)-E(U \rho U^\dagger)\,,
\end{equation}
where the average energy of the initial and final state is calculated with respect to the Hamiltonian $H$, and we have defined $E(\rho)=\Tr{H\rho}$. The optimal work extraction is achieved by maximizing the average work in Eq.~\eqref{eq.workave} over all the unitary cycles, i.e.,
\begin{equation}
\mathcal E (\rho) \equiv \max_{U} W(\rho,U) = W(\rho,U_E)\geq 0\,
\end{equation}
The optimal value $\mathcal E (\rho)$ is called ergotropy, and it is achieved by performing the optimal unitary cycle with unitary operator
\begin{equation}
U_E = \sum_{k=1}^d e^{i\phi_k} \ket{\epsilon_k}\bra{r_k}\,.
\end{equation}
The ergotropy is zero if and only if the initial state is passive, i.e., commutates with the Hamiltonian, $[\rho,H]=0$, and the populations with respect to the energy basis are sorted in decreasing order, $r_k=\bra{\epsilon_k} \rho\ket{\epsilon_k}$. The unitary cycle with operator $U_E$ lowers the average energy in the best way. However, the agent can be non-neutral to risk, in this case he will take in account the work fluctuations. 

\subsection{Expected utility hypothesis}\label{sec.utility}
Work can be thought of as a stochastic quantity $w$ affected by fluctuations.
For our purposes we aim to define a probability distribution of work, then we consider a general incoherent initial state in the basis of energy, i.e., a state such that $\rho=\Delta(\rho)$, where $\Delta(\rho)\equiv\sum_k \ket{\epsilon_k}\bra{\epsilon_k} \rho\ket{\epsilon_k}\bra{\epsilon_k}$ is the dephasing map. The initial state reads
\begin{equation}\label{eq.inistateinco}
\rho = \sum_{k=1}^d p_k \ket{\epsilon_k}\bra{\epsilon_k}\,,
\end{equation}
where $p_k$'s are related to $r_k$'s by a permutation $\pi_k$ of the indices, i.e., $r_k = p_{\pi_k}$ for all $k$. For instance, if the permutation is the identity $\pi_k=k$, we get $r_k=p_k$ and we can easily check that the state $\rho$ is passive.
For the unitary cycle with time evolution operator $U$, we define the extracted work as the random variable $w$  having the two-projective-measurement probability distribution~\cite{Talkner07,campisi11}
\begin{equation}
p(w,\rho,U)=\sum_{k,n} p_k |\bra{\epsilon_n}U\ket{\epsilon_k}|^2 \delta(w-\epsilon_k+\epsilon_n)\,.
\end{equation}
To introduce some notions of expected utility theory, we focus on an agent who must choose between two procedures that yield two different stochastic works $w_1$ and $w_2$ having certain probability distributions.
Thus, the works $w_\alpha$ is extracted in a procedure with initial state $\rho_\alpha$ and unitary time-evolution operator $U_\alpha$, for $\alpha=1,2$.
To give an example, we consider an agent who must choose between extracting a fixed work $w_{1}=w_{det}$ or flipping a coin and extracting a work $w_{2}=w_{head}$ if heads or nothing otherwise.
The utility function $u(w)$ quantifies the satisfaction gained from a choice, i.e., the risk aversion, so that the agent will choose the procedure extracting the work $w_1$ instead of $w_2$ if~\cite{bookmicroeco,bookmicroeco2}
\begin{equation}\label{eq.exp utility theo}
\langle u(w_1) \rangle > \langle u(w_2) \rangle\,,
\end{equation}
which represents the expected utility hypothesis.
Thus, in our example, an agent with the utility function $u(w)$ will choose to flip the coin if  $u(0)+u(w_{head})> 2 u(w_{det})$, is neutral to the choice if equality holds or will choose deterministic work $w_{det}$ otherwise.
It is easy to see that the inequality in Eq.~\eqref{eq.exp utility theo} remains unchanged if we perform an affine transformation on the utility function, i.e., the transformation $u(w) \mapsto a u(w) + b$, where $a$ is a positive variable. This means that the utility function is defined up to affine transformations. 
The work $w$ can be further characterized by the certainty equivalent, denoted with $w_{CE}$, defined such that
\begin{equation}\label{eq.wCEdef}
u(w_{CE}) = \langle u(w) \rangle\,.
\end{equation}
Thus, the certainty equivalent is obtained as the Kolmogorov-Nagumo average of the wealth, i.e., $w_{CE}=\langle w\rangle_{KN} \equiv u^{-1}(\langle u(w)\rangle)$, where $u(w)$ is a strictly monotonic function.
The meaning of the certainty equivalent becomes more clear if we consider as usual a strictly increasing utility function $u(w)$, so that Eq.~\eqref{eq.exp utility theo} is equivalent to $w^{CE}_1> w^{CE}_2$, where $w^{CE}_{1,2}$ is the certainty equivalent corresponding to the work $w_{1,2}$.
To understand in simple terms how the agent's risk aversion depends on the utility function $u(w)$, we start by noting that if $u(w)$ is a linear function, the certainty equivalent coincides with the average value, i.e., $w_{CE}=\langle w \rangle$. In this case, the agent prefers the procedure maximizing the average work and it is neutral to risk. {Instead, if $u(w)$ is a strictly increasing concave function, then the agent is averse to risk, since, by applying the Jensen's inequality to Eq.~\eqref{eq.wCEdef}, we get $w_{CE}<\langle w \rangle$. On the other hand, if $u(w)$ is a strictly increasing convex function, then the agent will prefer to risk since $w_{CE}>\langle w \rangle$.}
For instance, in our example, if $u(w)$ is a convex function, by applying Jensen's theorem we get the inequality $u(0) + u(w_{head})\geq 2 u(w_{head}/2)$.
Then, if $w_{head} > 2 w_{det}$ from the previous discussion the agent will flip the coin.
In summary, in terms of the certainty equivalent, we say that the agent is risk averse if $w_{CE}<\langle w\rangle$, risk neutral if $w_{CE}=\langle w\rangle$ and risk loving if $w_{CE}> \langle w\rangle$.
Furthermore, the utility function allows us to quantify how risk averse an agent is. For a utility function which is concave and strictly increasing, risk aversion can be measured with the Arrow-Pratt coefficient of absolute risk aversion defined as
\begin{equation}\label{eq. RA}
r_A(w) = -\frac{u''(w)}{u'(w)}\,,
\end{equation}
which is non-negative. It is clear that risk aversion depends on how much the utility function $u(w)$ is concave. Then, the  simpler quantifier of risk aversion should be the second derivative $u''(w)$. However, $u''(w)$ is not invariant under affine transformations, but we can work around this problem by dividing it by the first derivative $u'(w)$, which explains Eq.~\eqref{eq. RA}. More details can be found, e.g., in Refs.~\cite{bookmicroeco,bookmicroeco2}.

\section{Optimal work extraction}\label{sec.optimalinco}
Given the initial state $\rho$ in Eq.~\eqref{eq.inistateinco} and the initial Hamiltonian $H$ in Eq.~\eqref{eq.hami}, we aim to calculate the maximum over all the unitary cycles with unitary time-evolution $U$ of the average value of a utility function $u(w)$, i.e.,
\begin{equation}\label{eq.utilitymax}
\mathcal U (\rho) \equiv \max_{U} \int p(w,\rho,U) u(w) dw = \int p(w,\rho,U_u) u(w) dw\,.
\end{equation}
This equation defines the optimal expected utility $\mathcal U (\rho)$ and the optimal unitary cycle with unitary operator $U_u$. These are the main objects of investigation of this paper.
In particular, for a given unitary operator $U$, we get the average of the utility function
\begin{eqnarray}
\langle u(w) \rangle &=& \int p(w,\rho,U) u(w) dw \\
\label{eq.utidef}&=& \sum_{k,n}\bra{\epsilon_n}U\ket{\epsilon_k} p_k u(\epsilon_k-\epsilon_n)\bra{\epsilon_k}U^\dagger\ket{\epsilon_n}\,.
\end{eqnarray}
Then, we can get some advantage in the calculation of the maximum in Eq.~\eqref{eq.utilitymax} if $u(\epsilon_k-\epsilon_n)$ factorizes as $u(\epsilon_k-\epsilon_n)=u(\epsilon_k)u(-\epsilon_n)$. In this case, from Eq.~\eqref{eq.utidef} we get
\begin{eqnarray}
\langle u(w) \rangle &=& \sum_{k,n}\bra{\epsilon_n}U\ket{\epsilon_k} p_k u(\epsilon_k) \bra{\epsilon_k}U^\dagger\ket{\epsilon_n}u(-\epsilon_n) \\
&=& \Tr{U \rho u(H) U^\dagger u(-H)}\,.
\end{eqnarray}

\begin{figure}
[t!]
\centering
\includegraphics[width=0.95\columnwidth]{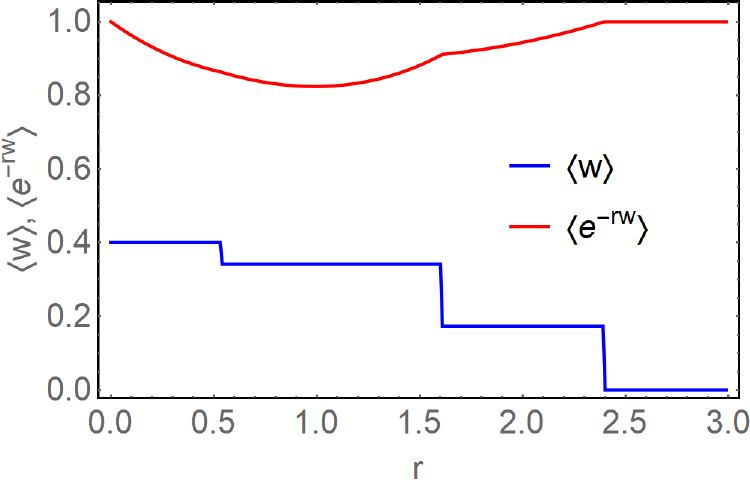}
\caption{ The average work $\langle w \rangle$ and $\langle e^{-r w}\rangle$  corresponding to the optimal unitary operator in the function of $r$ for a single battery with $d=3$. We put $p_1=0.1$, $p_2=0.4$, $p_3=0.5$, $\epsilon_1=0$, $\epsilon_2=0.579$ and $\epsilon_3=1$. We calculate the optimal unitary operator by selecting the unitary operator giving the minimum value of $\langle e^{-r w}\rangle$.
}
\label{fig:plotn1}
\end{figure}

A natural choice can be the exponential function $u(w)=-e^{-r w}/r$. To discuss further the characteristics of this choice, we focus on $r>0$, so that $u(w)$ is a concave function and thus the agent is averse to risk. In this case the agent aims to optimize the work fluctuations by minimizing the average  $\langle e^{-r w}\rangle$ over all the unitary cycles. To explain this, we note that if the probability distribution $p(w,\rho,U)$ is a Gaussian having variance $\sigma_w^2=\langle w^2\rangle - \langle w\rangle^2$, we get $\langle e^{-r w}\rangle = e^{-r \langle w \rangle + r^2 \sigma_w^2/2}$, then the agent wants to maximize the quantity $\langle w\rangle-r \sigma_w^2/2$. In simple terms, the agent prefers to reduce the spread along with getting maximum work. In general, minimizing $\langle e^{-r w}\rangle$ is equivalent to maximize the Kolmogorov-Nagumo average $\langle w \rangle_{KN}$, obtaining the maximum
\begin{equation}\label{eq.ECfirst}
\mathcal E_{CE}(\rho) = \max_U \langle w \rangle_{KN} = -\frac{1}{r}\min_U \ln \langle e^{-r w}\rangle\,.
\end{equation}
In our case, the Kolmogorov-Nagumo average can be expressed in terms of the cumulants series
\begin{equation}\label{eq.Kolmogorovave}
\langle w \rangle_{KN} = \sum_{n=1}^{\infty}\frac{(-r)^{n-1}}{n!}\kappa_n(w)\,,
\end{equation}
where $\kappa_n(w)$ is the n-th cumulant of the work, e.g., $\kappa_1(w) = \langle w\rangle$, $\kappa_2(w) = \sigma_w^2$, $\kappa_3(w) = \langle (w-\langle w\rangle)^3\rangle$, $\kappa_4(w) =\langle (w-\langle w\rangle)^4\rangle - 3 \sigma_w^4$, and so on. In particular, the average of Eq.~\eqref{eq.Kolmogorovave} has been recently discussed in Ref.~\cite{Morales23} when $w$ is replaced with the energy.
A further explanation is given by considering the Chernoff bound
\begin{equation}
\text{Pr}(w<x) \leq \langle e^{-rw}\rangle e^{rx}\,,
\end{equation}
where $\text{Pr}(w<x)$ is the probability to get a work smaller than a fixed amount $x$.
Thus, by minimizing $\langle e^{-rw}\rangle$, we minimize the upper bound of the probability $\text{Pr}(w<x)$ and so the left tail of the work probability distribution. In particular, this bound can be asymptotically saturated for a large number of quantum batteries for an optimal value of $r$, and $x<\langle w \rangle$, as we will see in Sec.~\ref{sec.ensemble}. Thus, in this case the agent maximizes the probability $\text{Pr}(w\geq x)=1-\text{Pr}(w<x)$ to get a work $w$ non-smaller than $x$.
Furthermore, we note that, for a given state $\rho$, work fluctuations are typically connected to the average work: If the agent wants to optimize work fluctuations somehow, e.g., minimizing $\langle e^{-rw}\rangle$, he can inevitably reduce the work extracted in average (see, e.g., Fig.~\ref{fig:plotn1}).
In the end, we note that, from the expected utility hypothesis, given a set of states, the agent prefers to extract work from the state with the largest value of $\mathcal E_{CE}$, where the fluctuations are minimal, although the corresponding average work, i.e., the ergotropy $\mathcal E$, can be lower than that of other states. Finally, a similar discussion can be done for $r<0$. In this case, the agent is risk loving and will prefer large fluctuations.

\subsection{Exponential utility function}\label{sec.exputi}
We focus on the following exponential utility function defined such that the absolute risk aversion, Eq.~\eqref{eq. RA}, is constant
\begin{equation}\label{eq.uteqxpo}
u(w) = \frac{1}{r}(1-e^{-rw})
\end{equation}
for $r\neq 0$, and $u(w)=w$ for $r=0$, which is a strictly increasing function. The agent is risk averse for $r>0$, risk neutral for $r=0$ and risk loving for $r<0$, and the absolute risk aversion of Eq.~\eqref{eq. RA} is constant and it is $r_A(w)=r$. For a linear $u(w)$, i.e., $r=0$, we get the ergotropy and $U_u=U_E$.
In order to calculate the maximum in Eq.~\eqref{eq.utilitymax} over all the unitary operators we note that:
\begin{theorem}\label{corollary}
In general, for any $r$ the stationary points are incoherent unitary operators expressed as
\begin{equation}\label{eq.uinco}
U_I = \sum_{k=1}^d e^{i\phi_k} \ket{\epsilon_k}\bra{\epsilon_{\pi_k}}\,,
\end{equation}
where $\pi_k$ is a permutation.
\end{theorem}
Thus, we have $d!$ non-trivial stationary points since the phases $\phi_k$ play a trivial role, i.e., do not appear in the expectation value of the utility function.

\begin{proof} To prove that the stationary points are of the form of Eq.~\eqref{eq.uinco}, we start by considering
\begin{equation}
\langle u(w) \rangle = \frac{1}{r}\left(1- \Tr{U \rho v(H) U^\dagger v(-H)}\right)\,,
\end{equation}
where we have define the function $v(w)=e^{-r w}$, so that $v(x+y)=v(x)v(y)$. We calculate the variation
\begin{equation}
\delta\langle u(w) \rangle \propto\Tr{\delta U \rho v(H) U^\dagger v(-H)}+\Tr{ U \rho v(H)\delta U^\dagger v(-H)}\,.
\end{equation}
We note that $\delta U= X U$ where $X$ is infinitesimal and such that $ X^\dagger=- X$. A variation over the set of the unitary operators can be obtained by considering a curve in the set, which has the Schr\"{o}dinger equation $\dot U = M U$, with $M^\dagger=-M$, where the derivative has been done with respect to the parameter $s$ of the curve, thus the variation is given by $\delta U = \dot U \delta s = X U$, where $X = M \delta s$.
We get
\begin{equation}
\delta\langle u(w) \rangle \propto \Tr{X[U \rho v(H) U^\dagger, v(-H)]}\,.
\end{equation}
By considering $X$ arbitrary, we get the stationary condition
\begin{equation}\label{eq.comm}
[U \rho v(H) U^\dagger, v(-H)]=0\,.
\end{equation}
Since $v(-H)$ and $\rho v(H) $ are diagonal operators in the basis of energy, $U \rho v(H) U^\dagger$ must be diagonal in the same basis if the commutator in Eq.~\eqref{eq.comm} is zero, which implies the stationary points in Eq.~\eqref{eq.uinco} and completes the proof.
\end{proof}

Then, Eq.~\eqref{eq.utilitymax} reduces to a maximum over all the $d!$ permutations $\pi_k$
\begin{equation}
\mathcal U (\rho) = \max_{\pi_k}\frac{1}{r}\left(1 - \sum_{n=1}^d p_{\pi_n}e^{-r( \epsilon_{\pi_n}-\epsilon_n)}\right)\,.
\end{equation}
To calculate the optimal value $\mathcal U (\rho)$, we define $y_k=e^{r \epsilon_k}/r$, so that $y_k < y_{k+1}$. Then, we sort  $\{p_{k}e^{-r \epsilon_{k}}\}$ in decreasing order, i.e., we choose the permutation $\pi_k$ such that $p_{\pi_k}e^{-r \epsilon_{\pi_k}}=u_k$ with $u_k\geq u_{k+1}$. From these considerations, we get our main result
\begin{equation}\label{eq.mainres}
\mathcal U (\rho) = \frac{1}{r}\left(1 - \sum_{k=1}^d u_k e^{r\epsilon_k}\right)\,.
\end{equation}
Concerning the optimal certainty equivalent, from Eq.~\eqref{eq.wCEdef} we get
\begin{equation}\label{eq.CE}
\mathcal E_{CE}(\rho) = u^{-1}(\mathcal U (\rho))= -\frac{1}{r}\ln\left(1-r\mathcal U (\rho)\right)\,,
\end{equation}
which is equivalent to Eq.~\eqref{eq.ECfirst} and  generalizes the ergotropy $\mathcal E(\rho)$ to an agent non-neutral to risk. Let us study these objects.
Operatively, the optimal values can be calculated by sorting the elements $\{p_{k}e^{-r \epsilon_{k}}\}$ in decreasing order achieving the sorted $u_k$. If for a given $r$ the elements are already sorted in decreasing order, i.e., $p_{k}e^{-r \epsilon_{k}}\geq p_{k+1}e^{-r \epsilon_{k+1}}$ or equivalently $u_k=p_{k}e^{-r \epsilon_{k}}$, the agent prefers to do nothing. In this case  $\mathcal{U}(\rho)=0$.
Interestingly, if there are at least two populations $p_{k}$ and $p_n$ different from zero, there is always some $r$ such that $u_k\neq p_{k}e^{-r \epsilon_{k}}$, and for this $r$ we get $\mathcal U (\rho)>0$, and the agent prefers to try to extract work instead to do nothing. To prove it, we consider $k>n$ and we search $r$ such that
\begin{equation}
p_n e^{-r \epsilon_n} < p_{k} e^{-r \epsilon_{k}}\,,
\end{equation}
which is satisfied for
\begin{equation}
r< \frac{1}{\epsilon_{n}-\epsilon_k}\ln \frac{p_n}{p_{k}}\,.
\end{equation}
Thus, we deduce that:
\begin{theorem}\label{theo2}
$\mathcal U (\rho)>0$ if and only if
\begin{equation}\label{eq.condition}
r < \max_{k,n| k>n} \frac{1}{\epsilon_n-\epsilon_k}\ln\frac{p_n}{p_k} = \frac{1}{\epsilon_{\bar {n}}-\epsilon_{\bar{k}}}\ln\frac{p_{\bar{n}}}{p_{\bar{k}}}\,.
\end{equation}
\end{theorem}
For instance, if the initial state is thermal $p_k=e^{-\beta \epsilon_k}/Z$, where $Z=\sum_k e^{-\beta \epsilon_k}$ and $\beta$ is the inverse temperature, then the condition in Eq.~\eqref{eq.condition} reduces to $r<-\beta$.
It results that $\mathcal{U}(\rho)\geq0$ and it is zero if and only if $p_{k+1}e^{-r \epsilon_{k+1}}\leq  p_{k}e^{-r \epsilon_{k}}$ for all $k$, or equivalently Eq.~\eqref{eq.condition} is not satisfied. States such that $p_{k+1}e^{-r \epsilon_{k+1}}\leq  p_{k}e^{-r \epsilon_{k}}$ for all $k$ are a generalization of the passive states achieved for $r=0$.
Concerning the optimal certainty equivalent in Eq.~\eqref{eq.CE}, we note that since $\mathcal E_{CE}(\rho)$ needs to be real, from Eq.~\eqref{eq.CE} it results the bound $\mathcal U (\rho)<1/r$ when $r>0$. In this case we get $\mathcal E_{CE}(\rho) \geq 0$ since the argument of the logarithm  is smaller than one. Furthermore, by noting that for $r=0$ we get the ergotropy and for $r<0$ the argument of the logarithm  is larger than one, we get that in general  $\mathcal E_{CE}(\rho) \geq 0$. Since $u(0)=0$, we get $\mathcal E_{CE}(\rho) = 0$ if and only if $\mathcal U(\rho) =0$. Moreover, since for $r>0$ we have $0\leq\mathcal U (\rho)<1/r$, we get $\mathcal U (\rho)\to 0$ as $r\to \infty$.
We note that $\mathcal U (\rho) $ and $\mathcal E_{CE}(\rho)$ can be also related to the Tsallis and R\'{e}nyi divergences defined as
\begin{eqnarray}
D^{\text{Tsallis}}_{\alpha}(p||q) &=&\frac{1}{\alpha-1}\left(\sum_j p^\alpha_j q^{1-\alpha}_j-1\right)\,, \\
D^{\text{R\'{e}nyi}}_{\alpha}(p||q) &=&\frac{1}{\alpha-1}\ln\left(\sum_j \frac{p^\alpha_j}{q^{\alpha-1}_j}\right)\,,
\end{eqnarray}
which for $\alpha>0$ are non-negative and equal to zero if and only if $p=q$, where $p=\{p_i\}$ and $q=\{q_i\}$ are two probability distributions such that $p_i\geq 0$, $\sum_i p_i = 1$, $q_i\geq 0$ and $\sum_i q_i = 1$. It is easy to see that there is a relation for the case of thermal populations $p_k=e^{-\beta \epsilon_k}/Z$. In this case, by defining $q_k = p_{\pi_k}$, from Eqs.~\eqref{eq.mainres} and~\eqref{eq.CE} we get
\begin{eqnarray}
\mathcal U(\rho) &=& -\beta D^{\text{Tsallis}}_{\alpha}(q||p) \,,\\
\mathcal E_{CE}(\rho) &=& -\beta D^{\text{R\'{e}nyi}}_{\alpha}(q||p)\,,
\end{eqnarray}
where $\alpha = 1+r/\beta$. In particular, for $\alpha>0$, namely $r>-\beta$, both the divergences are non-negative for any permutation $\pi_k$, which means that $p=q$ and the optimal permutation is $\pi_k=k$. This is in perfect agreement with the condition in Eq.~\eqref{eq.condition}.

To see how the optimal values $\mathcal U$ and $\mathcal E_{CE}$ differ from the egotropy, we start to define the characteristic function of work as $\chi(x,\rho,U)= \int e^{i x w} p(w,\rho,U)dw$ (alternatively also the definition $\chi(x,\rho,U)= \int e^{- x w} p(w,\rho,U)dw$ is typically used), which is the generating function of the work moments, and the generating function of work cumulants $g(x,\rho,U)=\ln \chi (x,\rho,U)$.
In general, the optimal values can be expressed as
\begin{equation}
\mathcal U (\rho) = \frac{1}{r}(1-\chi(ir,\rho,U_u))\,,\quad\mathcal E_{CE}(\rho) = -\frac{g(ir,\rho,U_u)}{r}\,.
\end{equation}
Thus, to achieve the optimal unitary cycle, we are minimizing $\chi(ir,\rho,U)/r$ over all the unitary cycles, instead of the average energy of the final state $E(U \rho U^\dagger)$ for the ergotropy. Since the optimal incoherent unitary changes by jumping from a permutation to a different one, we expect that there is a neighborhood of zero such that if $r\in(\delta_-,\delta_+)$ we get the optimal unitary operator $U_u=U_E$, where $\delta_-<0$ and $\delta_+>0$. Thus, for $r\in(\delta_-,\delta_+)$ we get
\begin{equation}
\mathcal U (\rho) = \frac{1}{r}(1-\chi(ir,\rho,U_E))\,,\quad\mathcal E_{CE}(\rho) = -\frac{g(ir,\rho,U_E)}{r}\,,
\end{equation}
from which, we have
\begin{eqnarray}\label{eq.usmallr}
\mathcal U (\rho) &=& \mathcal E(\rho) - \frac{r}{2}\langle w^2 \rangle_{E}+\mathcal O(r^2)\,,\\
\quad\mathcal E_{CE}(\rho) &=& \mathcal E(\rho) - \frac{r}{2}\left(\langle w^2 \rangle_{E}-\mathcal E^2(\rho)\right)+\mathcal O(r^2)\,,
\end{eqnarray}
where $\langle w^n \rangle_{E} = \int w^n p(w,\rho,U_E)dw$, e.g., $\langle w \rangle_E = \mathcal E(\rho)$.  Given two states $\rho$ and $\rho'$, by following the expected utility hypothesis an agent prefers the one with largest optimal value $\mathcal{U}$ (or equivalently $\mathcal E_{CE}$). Thus, if the agent is near to be neutral to risk, $r\in(\delta_-,\delta_+)$, the selection of the initial state is determined by the ergotropy and its higher moments (or cumulants).
In general, given two states $\rho$ and $\rho'$, it is important to compare the optimal expected utilities of the two states. To deduce a sufficient condition for the inequality $\mathcal{U}(\rho)\geq\mathcal{U}(\rho')$, we start by considering
\begin{eqnarray}
&&\mathcal{U}(\rho)-\mathcal{U}(\rho') = \frac{1}{r}\sum_{k=1}^d (u'_k-u_k)e^{r\epsilon_k}\\
&&=\sum_{k=1}^{d-1} \frac{e^{r\epsilon_{k+1}}-e^{r\epsilon_k}}{r} \sum_{j=1}^k (u_j-u_j')-\frac{e^{r \epsilon_d}}{r}\sum_{j=1}^d (u_j-u_j')\,,
\end{eqnarray}
where we used the summation by parts \cite{Allahverdyan2005}. All the parameters with apostrophe, e.g., $u'_k$, correspond to $\rho'$.
Since $\epsilon_k<\epsilon_{k+1}$, and $y(x)=e^{rx}/r$ is a strictly increasing function, we get $(e^{r\epsilon_{k+1}}-e^{r\epsilon_k})/r>0$, thus the weak majorization gives a sufficient condition, which is
\begin{eqnarray}\label{eq.compareweak}
\nonumber&&\{u_k\} \succ_w \{u'_k\} \text{ and }\left(r<0 \text{ or } \frac{1}{r}\sum_{j=1}^d (u_j-u_j')=0\right)\\
\nonumber &&\Rightarrow \mathcal{U}(\rho)\geq \mathcal{U}(\rho')\\
 &&\Rightarrow \mathcal{E}_{CE}(\rho)\geq \mathcal{E}_{CE}(\rho')\,,
\end{eqnarray}
where the weak majorization is defined such that $\{u_k\} \succ_w \{u'_k\}$ if and only if
\begin{equation}
\sum_{j=1}^k u_j\geq \sum_{j=1}^k u_j'\quad \text{for } k=1,\ldots,d\,.
\end{equation}
As $r\to 0$, we get that $\mathcal{U}(\rho)-\mathcal{U}(\rho')$ is equal to the ergotropy difference $\mathcal{E}(\rho)-\mathcal{E}(\rho')$ and the condition $\sum_{j=1}^d (u_j-u_j')/r=0$ is equal to $E(\rho) = E(\rho ')$, i.e., the two states have the same average energy.
Of course, also the majorization gives a sufficient condition for $r\neq 0$, which is
\begin{eqnarray}\label{eq.compare}
\{u_k\} \succ \{u'_k\} \Rightarrow \mathcal{U}(\rho)\geq \mathcal{U}(\rho')\Rightarrow \mathcal{E}_{CE}(\rho)\geq \mathcal{E}_{CE}(\rho')\,,
\end{eqnarray}
where the majorization is defined such that $\{u_k\} \succ \{u'_k\}$ if and only if
\begin{equation}
\sum_{j=1}^k u_j\geq \sum_{j=1}^k u_j'\quad \text{for } k=1,\ldots,d-1
\end{equation}
and $\sum_{k=1}^d u_k = \sum_{k=1}^d u'_k$.
This implies that both $\mathcal U(\rho)$ and $\mathcal E_{CE}(\rho)$ are Schur-concave functionals of $\{u_k\}$. Let us now give some examples.

\subsubsection{d=2}\label{sec.exputid2}

\begin{figure}
[t!]
\centering
\includegraphics[width=0.87\columnwidth]{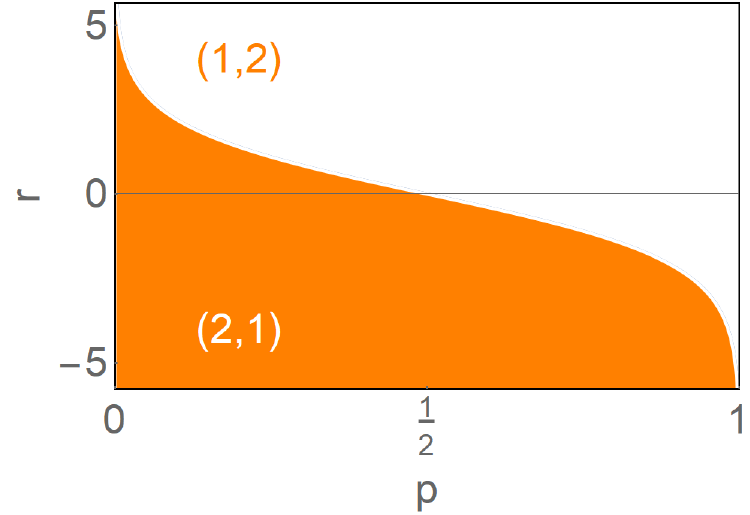}
\caption{ Optimal permutation in the plane $(p,r)$. In the orange region we get the NOT $(2,1)$, in the white region the identity $(1,2)$. The critical value $r=r^*$ is the bottom boundary of the white region.
}
\label{fig:plotd2}
\end{figure}

For $d=2$ we get a qubit, where the computational basis can be defined as $\ket{\epsilon_1}=\ket{0}$ and $\ket{\epsilon_2}=\ket{1}$. We have only two permutations which are the identity $\pi^{(I)}_k$, defined as $\pi^{(I)}_1=1$ and $\pi^{(I)}_2=2$, i.e., in the usual notation $(1,2)$, and the NOT gate $\pi^{(NOT)}_k$ defined as $\pi^{(NOT)}_1=2$ and $\pi^{(NOT)}_2=1$, i.e., in the usual notation $(2,1)$.
It is easy to see that the optimal $U_u$ is the identity for $r>r^*$, whereas we get the NOT permutation $(2,1)$ and $\mathcal U (\rho)>0$ for $r<r^*$ (see Fig.~\ref{fig:plotd2}). To find $r^*$ in the function of the populations, we require that the expected utility $\langle u(w) \rangle $ for $(2,1)$, which reads
\begin{equation}
\langle u(w) \rangle_{(2,1)} = \frac{1}{r}\left( 1- p_2 e^{- r (\epsilon_2-\epsilon_1)} -p_1 e^{- r (\epsilon_1-\epsilon_2)}   \right)\,,
\end{equation}
is zero. From this condition we get
\begin{equation}\label{eq.rstard2}
r^* = \frac{1}{\epsilon} \ln\frac{1-p}{p}\,,
\end{equation}
where $\epsilon=\epsilon_2-\epsilon_1$ and $p=p_1$.
We note that for $r=0$, the identity permutation $(1,2)$ is selected for $p_1>1/2$ (passive state), the NOT permutation $(2,1)$ for $p_1<1/2$ (active state), in agreement with the ergotropy, that is nonzero only if the state is active. For an initial active state such that $p_1<1/2$, we get $r^*>0$, so that a risk averse agent for $0<r<r^*$ prefers to gamble and to perform the cycle with NOT permutation, since the state is active and average work can be extracted. For an initial passive state $p_1>1/2$, we get $r^*<0$, so that a risk loving agent prefers to do nothing (i.e., the identity) for $0>r>r^*$ instead to gamble performing the NOT permutation $(2,1)$. We can have a deterministic work extraction, i.e., the probability distribution is a Dirac delta,  when $p_1=1$ or $p_2=1$. When $p_1=1$ we get the ground-state $\rho=\ket{\epsilon_1}\bra{\epsilon_1}$, and the identity is preferred for any $r$. Instead for $p_2=1$ we get the excited-state $\rho=\ket{\epsilon_2}\bra{\epsilon_2}$, and the NOT is preferred for any $r$. This is because in both cases we obtain a deterministic protocol for extracting the work.

\subsubsection{d=3}\label{sec.exputid3}
\begin{figure}
[t!]
\centering
\includegraphics[width=0.487\columnwidth]{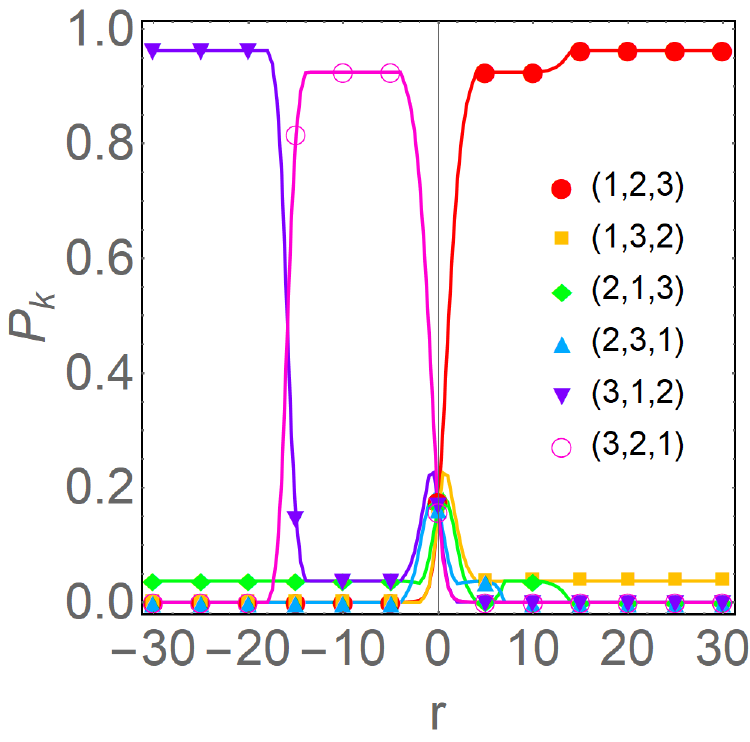}\includegraphics[width=0.487\columnwidth]{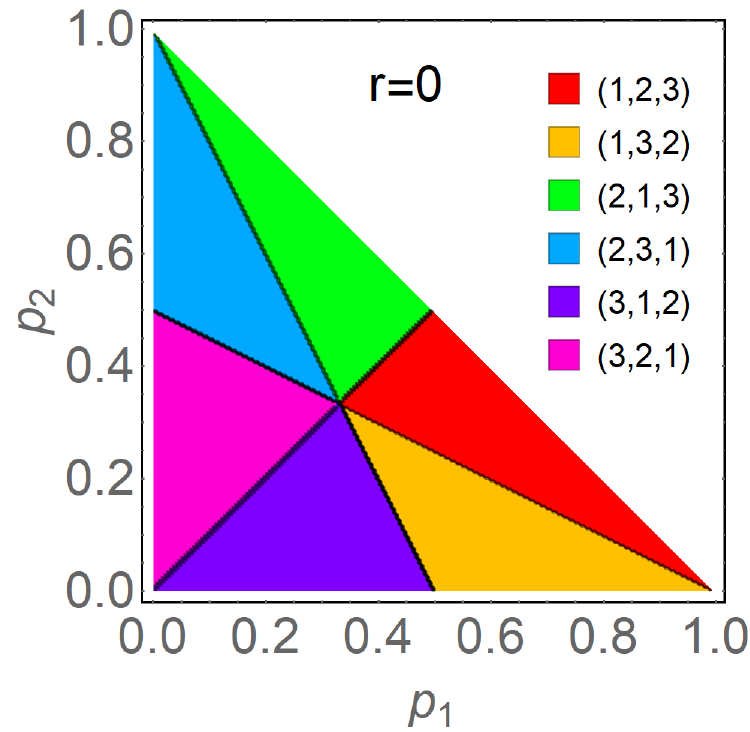}\\
\includegraphics[width=0.487\columnwidth]{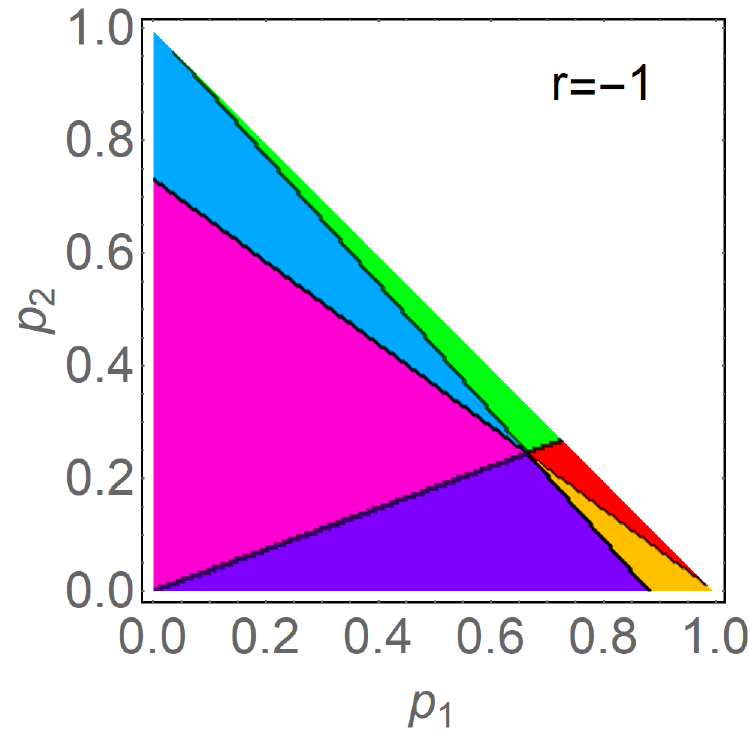}\includegraphics[width=0.487\columnwidth]{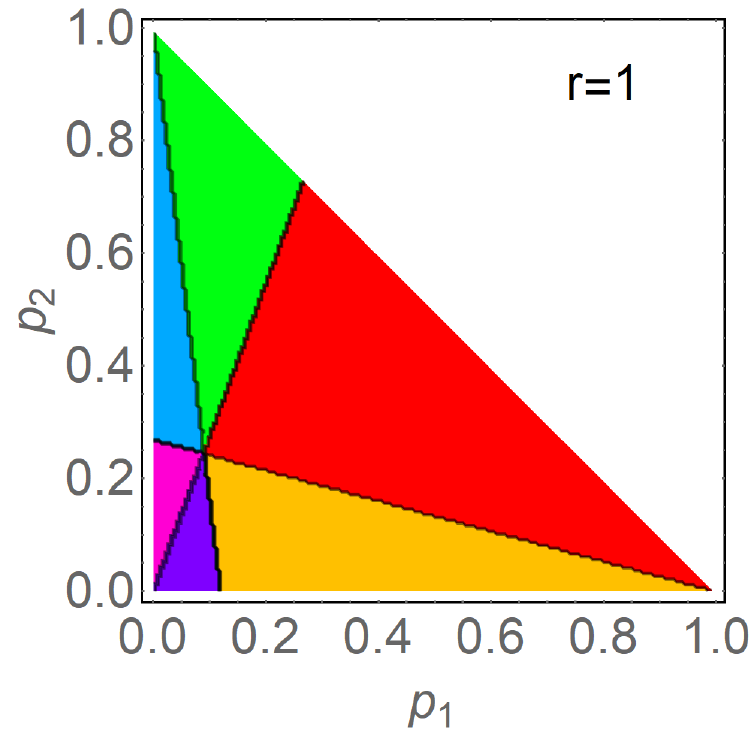}
\caption{In the top left panel, we plot the probability $P_k$ in the function of $r$. $P_k$ is the probability to get a permutation $k$, where $k=(1,2,3),(1,3,2),\ldots, (3,2,1)$, by choosing randomly $p_1$ and $p_2$, e.g., by flipping a coin. In detail, $P_k$ is calculated as the frequency to get a certain permutation in the plane $(p_1,p_2)$. In the remaining panel, the optimal permutation in the plane $(p_1,p_2)$ for different value of $r$. In detail, we put $r=0$ (top right panel), $r=-1$ (bottom left panel) and $r=1$ (bottom right panel). We consider the energies $\epsilon_k=k$.
}
\label{fig:plotd3}
\end{figure}

For $d=3$, i.e., a qutrit, we have $3!=6$ permutations. We focus on an active state (the ergotropy is nonzero) with $p_{k+1}\geq p_{k}$ for all $k$. For $r= 0$, we get the ergotropic permutation $(3,2,1)$, and the average utility is equal to the ergotropy and it is nonzero. Let us see how permutation are selected as the risk aversion change. For $r\neq 0$, for this permutation we get
\begin{equation}
\langle u(w) \rangle_{(3,2,1)} = \frac{1}{r}\left(1 -  p_{3}e^{-r( \epsilon_{3}-\epsilon_1)}-p_{2}-p_{1}e^{-r( \epsilon_{1}-\epsilon_3)}\right)\,.
\end{equation}
As $r\to - \infty$, we get
\begin{equation}
\langle u(w) \rangle_{(3,2,1)} \sim \frac{1}{r}\left(1 -p_{2}-p_{3}e^{-r( \epsilon_{3}-\epsilon_1)}\right)\,.
\end{equation}
On the other hand, for the permutation $(3,1,2)$ we get
\begin{equation}
\langle u(w) \rangle_{(3,1,2)} = \frac{1}{r}\left(1 -  p_{3}e^{-r( \epsilon_{3}-\epsilon_1)}-p_{1}e^{-r( \epsilon_{1}-\epsilon_2)}-p_{2}e^{-r( \epsilon_{2}-\epsilon_3)}\right)\,.
\end{equation}
As $r\to-\infty$ we get
\begin{equation}
\langle u(w) \rangle_{(3,1,2)} \sim \frac{1}{r}\left(1 -p_{3}e^{-r( \epsilon_{3}-\epsilon_1)}\right)\,.
\end{equation}
Thus, since $\epsilon_{3}-\epsilon_1$ is the largest difference and for $p_3\neq 0$ appears only in the expected utility with permutations $(3,2,1)$ and $(3,1,2)$, as $r$ decreases from zero to $-\infty$, if $p_2\neq 0$ the optimal permutation jumps from $(3,2,1)$ to $(3,1,2)$.
Conversely, for $r\to \infty$ all the utility expectation values are negative except the identity that is zero, thus the optimal permutation is the identity $(1,2,3)$. However, for intermediate values of $r$, between the permutations $(3,2,1)$ and $(1,2,3)$ there may be other permutations, depending on the values of $p_k$ and the values of energies $\epsilon_k$.
For other states, such that $p_{k+1}\geq p_k$ is not satisfied for some $k$, the ergotropic permutation is not $(3,2,1)$, but a similar behavior occurs (see Fig.~\ref{fig:plotd3}).

\subsubsection{d=2+2}\label{sec.exputid2+2}
We consider a bipartite system of two qubits. The Hilbert space of a qubit has the computation basis $\{\ket{0},\ket{1}\}$. The Hamiltonian of a qubit is $H_{qubit}=\epsilon \ket{1}\bra{1}$, so that the Hamiltonian of the total system is $H=I\otimes H_{qubit}+H_{qubit}\otimes I$, giving the energies $\epsilon_1=0$, $\epsilon_2=\epsilon_3=\epsilon$ and $\epsilon_4=2 \epsilon$, and the eigenstates $\ket{\epsilon_1}=\ket{00}$, $\ket{\epsilon_2}=\ket{01}$, $\ket{\epsilon_3}=\ket{10}$ and $\ket{\epsilon_4}=\ket{11}$.
We start to consider the initial state $\rho=\rho_{qubit}^{\otimes 2}$, where $\rho_{qubit}=p \ket{0}\bra{0}+(1-p)\ket{1}\bra{1}$. Thus, the populations of the total system are $p_1= p^2$, $p_2=p_3=p(1-p)$ and $p_4=(1-p)^2$.
In this case, it is easy to see that the ergotropy $\mathcal E (\rho)$ can be extracted with an incoherent unitary operator $U_E= U_{E,qubit}^{\otimes 2}$ without generating correlations in the final state $U_E \rho U_E^\dagger$. The local incoherent unitary operator $U_{E,qubit}$ is the ergotropic one for $d=2$. We have $U_{E,qubit}=I$ if the state $\rho_{qubit}$ is passive, i.e., for $p\geq 1/2$, whereas $U_{E,qubit}=U_{NOT}\equiv\ket{0}\bra{1}+\ket{1}\bra{0}$  for $p<1/2$.
For $r\neq 0$, we find that two permutations are essentially involved: the identity $\pi^{(I)}_k=k$, i.e., the permutation $(1,2,3,4)$, and the NOT permutation $\pi^{(NOT)}_k$, such that $\pi^{(NOT)}_1=4$, $\pi^{(NOT)}_2=3$, $\pi^{(NOT)}_3=2$, $\pi^{(NOT)}_4=1$,  i.e., the permutation $(4,3,2,1)$. Then, we have the optimal unitary operator $U_u=U_{qubit}^{\otimes 2}$, where $U_{qubit}=I$ if $r\geq r^*$, whereas $U_{qubit}=U_{NOT}$ if $r<r^*$, where $r^*$ is again the one in Eq.~\eqref{eq.rstard2}.
To prove it, we consider the permutation $(4,3,2,1)$ which gives
\begin{equation}
\langle u(w) \rangle_{(4,3,2,1)} = \frac{1}{r}\left( 1- p_4 e^{- 2 r\epsilon } -p_3 -p_2 -p_1 e^{2 r \epsilon}   \right)\,,
\end{equation}
so that $r^*$ is solution of $\langle u(w) \rangle_{(4,3,2,1)}=0$ and we get  Eq.~\eqref{eq.rstard2}.

\subsubsection{Initial correlations}
To study the role of initial correlations, we label the two qubits of Sec.~\ref{sec.exputid2+2} with $A$ and $B$, and we consider an initial state showing correlations between the two parties $A$ and $B$. Thus, we consider an initial incoherent state in Eq.~\eqref{eq.inistateinco}, such that
$\rho \neq \pi_\rho$, where $\pi_\rho=\rho_A \otimes \rho_B$. The reduced states are given by calculating the partial trace, getting $\rho_A = \ParTr{B}{\rho}$ and $\rho_B = \ParTr{A}{\rho}$.
The mutual information $I_{A:B}(\rho)= S(\pi_\rho)-S(\rho)=S(\rho_A)+S(\rho_B)-S(\rho)$ is nonzero, so that there are only classical correlations, since the initial state $\rho$ is classical-classical. In detail,  $S(\rho)=-\Tr{\rho \ln \rho}$ is the von Neumann entropy.
We find that the optimal permutation corresponding to the initial state $\rho$ can be different from the one corresponding to the initial state $\rho'=\pi_\rho$. Furthermore, $u_k$ does not necessarily majorize $u'_k$, and there are states such that $\mathcal U (\rho)<\mathcal U (\pi_\rho)$ for $r<r_I$, whereas $\mathcal U (\rho)>\mathcal U (\pi_\rho)$ for $r>r_{I}$. For instance, for the state with populations $p_1=0.33$, $p_2=0.34$, $p_3=0.22$ and $p_4=0.11$, we get $r_I\approx-0.8$. In particular, for $r=0$, we get $\delta_T(\rho)\equiv\mathcal E(\rho) - \mathcal E (\pi_\rho)\geq 0$ for two qubits (see Ref.~\cite{Francica22ergo}), which suggests that $r_I$ is nonpositive. Then, for certain initial states, a very risk loving agent tends to prefer the uncorrelated state $\pi_\rho$ instead of the correlated state $\rho$, i.e., the agent prefers to discharge the correlations between the two qubits $A$ and $B$.

\subsubsection{Ensemble of quantum batteries}\label{sec.ensemble}
As a last non-trivial example, we consider an ensemble of $n$ quantum batteries, where every battery has the Hamiltonian $H$ in Eq.~\eqref{eq.hami} and it is prepared in the state $\rho$.
We recall that a state $\rho$ is called completely passive if $\rho^{\otimes n}$ is passive for any number $n$ of batteries. One can show that a state $\rho$ is completely passive if it is equal to the Gibbs state $\rho_\beta \equiv e^{-\beta H}/Z$, where $Z=\Tr{e^{-\beta H}}$ and $\beta$ is solution of $S(\rho_\beta)=S(\rho)$, where $S(\rho)$ is the von Neumann entropy. This can be understood by noting that $\mathcal E ( \rho^{\otimes n}) \leq n (E(\rho)- E (\rho_\beta))$ for any $n$. In particular, it is possible to perform a unitary cycle such that
\begin{equation}\label{eq.Alicki-Fannes}
\mathcal E ( \rho^{\otimes n}) \sim n (E(\rho)- E (\rho_\beta))
\end{equation}
as $n\to \infty$, which optimizes the average work extracted (see Ref.~\cite{Alicki13}).
Here we aim to calculate $\mathcal E_{CE}(\rho^{\otimes n})$ for the most interesting case $r>0$ and thus for a risk averse agent, which tries to minimize the work fluctuations. Thus, the agent aims to minimize
\begin{equation}\label{eq.trace}
\langle e^{-r w}\rangle =  \Tr{\rho^{\otimes n}\left( e^{-rH}\right)^{\otimes n} U^\dagger \left(e^{r H}\right)^{\otimes n} U}
\end{equation}
over all the unitary operators $U$.
We calculate the trace in Eq.~\eqref{eq.trace} with respect to the basis of the energy eigenstates $\ket{i_1, \ldots , i_d}\equiv \ket{\{i_k\}}= \otimes_{k=1}^n \ket{\epsilon_{i_k}}$ of the $n$ batteries system. Given a state $\ket{\{i_k\}}$, we define $n_i$ as the number of the indices $i_k$ that are equal to $i$. Then, the energy of this state is $\epsilon_{n_1 \ldots n_d}\equiv \epsilon_{\{n_i\}}=\sum_{i=1}^d n_i \epsilon_i$, so that its value is specified by the occupation numbers set $\{n_i\}$.
The Hilbert space of the $n$ batteries system can be expressed as the direct sum
\begin{equation}
\mathcal H = \bigoplus_{n_1,\ldots,n_d | \sum_i n_i=n}\mathcal H_{n_1\ldots n_d}\,,
\end{equation}
where  the subspace $\mathcal H_{n_1 \ldots n_d}$ has dimension
\begin{equation}
W_{n_1 \ldots n_d} = \frac{n!}{n_1!\cdots n_d!}\,.
\end{equation}
Namely, $W_{n_1 \ldots n_d}$ is calculated as the number of states $\ket{\{i_k\}}$ having occupation numbers $\{n_i\}$.

In general the average $\langle e^{-r w}\rangle$ is minimum when $U$ is an incoherent unitary operator, which performs a permutation of the states $\ket{\{i_k\}}$. This incoherent unitary operator, given an initial state $\ket{\{i_k\}}\in \mathcal H_{\{n_i\}}$ with the initial occupation numbers $\{n_i\}$, leads to a final state $\ket{\{\bar i_k\}}\in \mathcal H_{\{\bar n_i\}}$ with the final occupation numbers $\{\bar n_i\}$.
We can write this incoherent unitary operator as
\begin{equation}
U = \bigoplus_{n_1,\ldots,n_d | \sum_i n_i = n} U_{n_1\ldots n_d}\,,
\end{equation}
where $U_{n_1\ldots n_d}: \mathcal H_{n_1\ldots n_d}\to \mathcal H'_{n_1\ldots n_d} $. In particular, the subspace $\mathcal H '_{n_1\ldots n_d}$ has dimension $W_{\{n_i\}}$, and in general we can define the number
\begin{equation}
W_{\bar n_1\ldots \bar n_d|n_1 \ldots n_d} = \sum_{\ket{\{i_k\}}\in \mathcal H_{\{n_i\}}}\sum_{\ket{\{\bar{i}_k\}}\in \mathcal H_{\{\bar{n}_i\}}} \abs{\bra{\{\bar{i}_k\}}U_{\{n_i\}}\ket{\{i_k\}}}^2\,,
\end{equation}
where $\abs{\bra{\{\bar{i}_k\}}U_{\{n_i\}}\ket{\{i_k\}}}$ is zero or one, since $U$ is an incoherent unitary operator. Thus, we note that
\begin{equation}
\sum_{\bar n_1 \ldots \bar n_d |\sum_i \bar n_i = n} W_{\bar n_1\ldots \bar n_d|n_1 \ldots n_d} = W_{n_1 \ldots n_d}\,,
\end{equation}
from which we define the conditional probability
\begin{equation}
p_{\bar n_1\ldots \bar n_d|n_1 \ldots n_d} = \frac{ W_{\bar n_1\ldots \bar n_d|n_1 \ldots n_d} }{ W_{n_1 \ldots n_d} }\,.
\end{equation}
To evaluate the trace in Eq.~\eqref{eq.trace}, we note that for the state $\rho^{\otimes n}$ we get the occupation numbers $\{n_i\}$ (and so the energy $\epsilon_{n_1 \ldots n_d}$) with probability
\begin{equation}\label{eq.probene}
p_{n_1 \ldots n_d} = W_{n_1 \ldots n_d} p_1^{n_1}\cdots p_d^{n_d}\,.
\end{equation}
Thus, from Eq.~\eqref{eq.trace}, we get the expectation value
\begin{eqnarray}\label{eq.exp-rw}
\nonumber\langle e^{-rw}\rangle &=& \sum_{n_1\ldots n_d|\sum_i n_i=n} \sum_{\bar n_1 \ldots \bar n_d |\sum_i \bar n_i = n} p_{n_1\ldots n_d}p_{\bar n_1\ldots \bar n_d|n_1 \ldots n_d}\\
&&\times  e^{-r\sum_i(n_i-\bar n_i)\epsilon_i}\,.
\end{eqnarray}
The optimal $U$, i.e., the optimal conditional probabilities $p_{\bar n_1\ldots \bar n_d|n_1 \ldots n_d}$ are such that Eq.~\eqref{eq.exp-rw} is minimum. We note that, for a  given set $\{n_i\}$, determining $p_{\bar n_1\ldots \bar n_d|n_1 \ldots n_d}$ for all $\{\bar n_i\}$ is equivalent to determinate the operator $U_{\{n_i\}}$ up to irrelevant phases, since $U$ is an incoherent unitary operator.

To perform our calculations, we focus on a large number $n$ of batteries. Then, by defining $\tilde p_i = n_i/n$, we get
\begin{equation}
W_{n_1\ldots n_d} \sim e^{n H(\tilde p)}
\end{equation}
as $n\to \infty$, where $H(\tilde p)= -\sum_i \tilde p_i \ln \tilde p_i$ is the Shannon entropy of the distribution probability $\tilde p=\{\tilde p_i\}$. Thus, the probability in Eq.~\eqref{eq.probene} reads
\begin{equation}
p_{n_1 \ldots n_d} \sim e^{- n D(\tilde p || p)}\,,
\end{equation}
where $D(\tilde p || p) = -H(\tilde p) - \sum_i \tilde p_i \ln p_i$ is the Kullback–Leibler divergence, and we have defined the probability distribution $p =\{p_i\}$. Furthermore, without loss of generality we assume
\begin{equation}
p_{\bar n_1 \ldots \bar n_d|n_1 \ldots n_d} \sim e^{- n D(\bar p)}
\end{equation}
for a certain function $D(\bar p)\geq 0$ of the probability distribution $\bar p = \{\bar p_i\}$, with $\bar p_i = \bar n_i/n$.
To minimize Eq.~\eqref{eq.exp-rw} we can proceed similarly to the maximization of average work given in Appendix~\ref{app.AlickiFannes}, resulting in Eq.~\eqref{eq.Alicki-Fannes}. As $n\to\infty$ we get
\begin{eqnarray}
\nonumber \langle e^{-rw}\rangle &\sim& \sum_{n_1\ldots n_d|\sum_i n_i=n} \sum_{\bar n_1 \ldots \bar n_d |\sum_i \bar n_i = n}e^{-nD(\tilde p||p)}e^{-n D(\bar{p})}\\
&&\times e^{-nr\sum_i(\tilde p_i-\bar p_i)\epsilon_i}\,.
\end{eqnarray}
We can obtain a lower bound of this expression by considering that, for a given set $\{n_i=n \tilde p_i\}$,
\begin{equation}
 \sum_{\bar n_1 \ldots \bar n_d |\sum_i \bar n_i = n} p_{\bar n_1\ldots \bar n_d|n_1 \ldots n_d} \sum_i\bar n_i\epsilon_i  \sim n \sum_i p'_i\epsilon_i\,,
\end{equation}
where $\{n'_i=n p'_i\}$ with $H(p')=H(\tilde p)$, which is Eq.~\eqref{eq.001} proven in Appendix~\ref{app.AlickiFannes}. Thus
\begin{equation}
 \exp\left\{r\sum_{\bar n_1 \ldots \bar n_d |\sum_i \bar n_i = n} p_{\bar n_1\ldots \bar n_d|n_1 \ldots n_d} \sum_i\bar n_i\epsilon_i\right\}  \sim  e^{r\sum_i n'_i\epsilon_i}\,.
\end{equation}
Since $h(x)=e^{rx}$ is a convex function, we can use the Jensen's inequality $\langle h(x)\rangle\geq h(\langle x\rangle)$, from which
\begin{equation}
 \sum_{\bar n_1 \ldots \bar n_d |\sum_i \bar n_i = n} p_{\bar n_1\ldots \bar n_d|n_1 \ldots n_d}e^{r \sum_i\bar n_i\epsilon_i}  \geq   e^{r\sum_i n'_i\epsilon_i}\,.
\end{equation}
Then, it results the lower bound
\begin{equation}\label{eq.bound01}
\langle e^{-rw}\rangle \geq \sum_{n_1\ldots n_d|\sum_i n_i=n} e^{-nD(\tilde p||p)} e^{-nr\sum_i(\tilde p_i- p'_i)\epsilon_i}\,.
\end{equation}
As for the maximization of the average work (see Appendix~\ref{app.AlickiFannes}), only a distribution $\tilde p=\tilde p^*$ dominates in the sum, which is the one that makes stationary the Lagrangian $\tilde F[\tilde p]= f[\tilde p] + \mu \sum_i \tilde p_i$, where
\begin{equation}\label{eq.fINCO}
f[\tilde p]=D(\tilde p||p)+r\sum_i(\tilde p_i- {p'_i}^*)\epsilon_i
\end{equation}
and the choice $p'={p'}^*$ makes $\langle e^{-rw}\rangle$ minimal.
By requiring $\delta \tilde F[\tilde p] = 0$, we get that the stationary point $\tilde p=\tilde p^*$ satisfies the equations
\begin{equation}\label{eq.tildep02}
\ln \tilde p_i - \ln p_i + 1 + r\epsilon_i - r \sum_{j} \epsilon_j  \frac{\partial  {p'_j}^*}{\partial \tilde p_i} + \mu=0\,.
\end{equation}
Thus, from Eq.~\eqref{eq.bound01} we get the bound
\begin{equation}\label{eq.bound02}
\langle e^{-rw}\rangle \geq e^{-n f[\tilde p^*]}\,,
\end{equation}
where equality in Eq.~\eqref{eq.bound02} asymptotically holds for the conditional probabilities of the form in Eq.~\eqref{eq.satuprob}, then the bound in Eq.~\eqref{eq.bound02} can be saturated and the right side gives the minimum value of $\langle e^{-rw}\rangle$. The optimal unitary operator $U$ is such that $U_{\{n_i\}}: \mathcal H_{\{n_i\}}\to \mathcal H_{\{n'_i\}} $ for $\{n_i=n\tilde p^*_i\}$.
To determinate ${p'}^*$, we start by noting that $\tilde p^*_i$  depends only on ${p'_i}^*$ through the partial derivatives $\partial {p'_j}^*/\partial \tilde p^*_i$, with $j=1,\ldots, d$, then it is easy to see that $\partial \tilde p^*_i/\partial {p'_j}^* = 0$ for all $i,j$. The optimal $p'={p'}^*$ makes stationary the Lagrangian $F'[p']=f[\tilde p^*]+\lambda (H(p')-H(\tilde p^*))+\nu \sum p'_i$ (where in $f[\tilde p^*]$ we replaced ${p'_i}^*$ with $p'_i$), from which we get the equations
\begin{equation}
r \epsilon_i + \lambda (\ln p'_i +1) + \nu = 0\,,
\end{equation}
giving
\begin{equation}\label{eq.pprime}
{p'_i}^* = \frac{e^{-\beta \epsilon_i}}{Z}\,,
\end{equation}
with $Z = \sum_i e^{-\beta \epsilon_i}$ and $\beta$ solution of $H({p'}^*)=H(\tilde p^*)$. Thus, the optimal certainty equivalent reads
\begin{equation}\label{eq.ECEnbatt}
\mathcal E_{CE}(\rho^{\otimes n}) \sim \frac{n f[\tilde p^*]}{r}\,,
\end{equation}
which is checked for $d=3$ in Fig.~\ref{fig:plot-bound}. An approximated expression for small $r$ is given in Appendix~\ref{app.smallr}.
\begin{figure}
[t!]
\centering
\includegraphics[width=0.95\columnwidth]{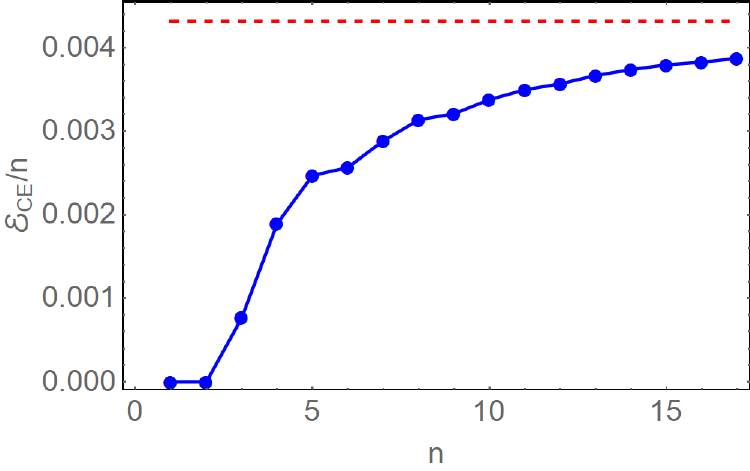}
\caption{ The optimal certainty equivalent $\mathcal E_{CE}(\rho^{\otimes n})$ versus the number $n$ of batteries (blue points). We put $r=1$, $p_1=0.538$, $p_2=0.237$, $p_3=0.225$, $\epsilon_1=0$, $\epsilon_2=0.579$ and $\epsilon_3=1$. For the asymptotic value in Eq.~\eqref{eq.ECEnbatt}(red dashed line) we get $f[\tilde p^*]\approx 0.0043$.
}
\label{fig:plot-bound}
\end{figure}

It is important to note that for $r\neq0$ we get $\tilde p^*\neq p$, thus we can minimize and maximize simultaneously the asymptotic values of $\langle e^{-rw}\rangle$ and $\langle w \rangle$, respectively, differently from the result shown in Fig.~\ref{fig:plotn1} for a single battery. This because the maximization of the asymptotic value of $\langle w \rangle$ fixes only the operator $U_{\{n_i\}}$ for typical occupation numbers $\{n_i=n p_i\}$. Then, for $r\neq 0$ we are free to choose $U_{\{n_i\}}$ with $\{n_i=n \tilde p^*_i\}$ in order to get the minimum asymptotic value of $\langle e^{-rw}\rangle$.
Furthermore, for large $n$, the Chernoff bound is saturated as usual, and we get
\begin{equation}\label{eq.largedevi}
\text{Pr}(w<x)\sim e^{-nI(x)} = e^{-n\sup_{r>0}(f[\tilde p^*]-rx/n)}\,,
\end{equation}
where $I(x)$ is the so-called rate function. We note that Eq.~\eqref{eq.largedevi} defines a relation $x(r)$ between $x$ and the optimal $r$. This explains how calculating $\mathcal E_{CE}$ for a given risk aversion $r$ is equivalent to minimize the probability $\text{Pr}(w<x)$ for $x=x(r)$, as $n\to \infty$.
In the end, as for the case $n=1$, we expect that there is a neighborhood of zero such that if $r\in(\delta_-,\delta_+)$ we get the optimal unitary operator $U_u=U_E$. In this case, for $r\in (\delta_-,\delta_+)$, we get the generating functions $\chi(ir,\rho^{\otimes n},U_E)\sim e^{-n f[\tilde p^*]}$ and $g(ir,\rho^{\otimes n},U_E)\sim -n f[\tilde p^*]$, e.g., the ergotropy cumulants can be calculated as $\kappa_{E,m}(w) \sim n (-1)^{m+1}\partial^m_r f[\tilde p^*]|_{r=0}$, where $\kappa_{E,1}(w)=\mathcal E(\rho^{\otimes n})$, $\kappa_{E,2}(w)= \langle w^2\rangle_E - \mathcal E^2(\rho^{\otimes n}) $, and so on. For instance, the first two cumulants can be obtained from the small $r$ expressions given in Appendix~\ref{app.smallr}.

\subsection{Arbitrary utility function}\label{sec.arbiuti}
For an arbitrary utility function we get Eq.~\eqref{eq.utidef}.
By considering $\delta U = X U$, we get the variation
\begin{eqnarray}\label{eq.arbiinco1}
\nonumber \delta\langle u(w) \rangle &=&  \sum_{k,n}\left(\bra{\epsilon_n}XU\ket{\epsilon_k} \bra{\epsilon_k}U^\dagger\ket{\epsilon_n}-\bra{\epsilon_n}U\ket{\epsilon_k}\bra{\epsilon_k}U^\dagger X\ket{\epsilon_n}\right)\\
&&\times p_k u(\epsilon_k-\epsilon_n)\,.
\end{eqnarray}
It is easy to see that the variation $\delta\langle u(w) \rangle$ is zero if $U=U_I$ where $U_I$ is an incoherent unitary operator of the form in Eq.~\eqref{eq.uinco}. In principle, we can also have non-incoherent unitary operators as stationary points.
If  the maximum of $\langle u(w) \rangle $ over all the unitary cycles is achieved for an incoherent unitary operator of the form of Eq.~\eqref{eq.uinco}, we get
\begin{equation}\label{eq.arbiinco}
\mathcal U (\rho) = \max_{\pi_k} \sum_{k=1}^d p_{\pi_k} u(\epsilon_{\pi_k}-\epsilon_k)\,.
\end{equation}
We find that Theorem~\ref{corollary} results to be a corollary of the following theorem:
\begin{theorem}
If the utility function $u(w)$ is a monotonic function, the stationary points are incoherent unitary operators of the form of Eq.~\eqref{eq.uinco}, from which we obtain Eq.~\eqref{eq.arbiinco}.
\end{theorem}
\begin{proof}
To prove it, we  write Eq.~\eqref{eq.arbiinco1} as
\begin{eqnarray}
\nonumber\delta \langle u(w)\rangle &=& \sum_{k,n,j} \bra{\epsilon_n}X\ket{\epsilon_j}\bra{\epsilon_j}U \ket{\epsilon_k}\bra{\epsilon_k}U^\dagger\ket{\epsilon_n}\\
&&\times p_k(u(\epsilon_k-\epsilon_n)-u(\epsilon_k-\epsilon_j))\,.
\end{eqnarray}
By considering $\bra{\epsilon_n}X\ket{\epsilon_j}$ arbitrary, from $\delta \langle u(w)\rangle=0$ we get
\begin{equation}\label{eq.arbiinco2}
\sum_{k=1}^d\bra{\epsilon_j}U \ket{\epsilon_k}\bra{\epsilon_k}U^\dagger\ket{\epsilon_n}p_k(u(\epsilon_k-\epsilon_n)-u(\epsilon_k-\epsilon_j))=0
\end{equation}
for all $n$ and $j$. The $d$ equations for $n=j$ are automatically satisfied. The equations for $j<n$ can be achieved from the equations $j>n$ by complex conjugation. Thus we focus only on the equations for $j>n$. If $u(w)$ is a monotonic function (on the support of the work), e.g., $u(x)>u(y)$ for any $x>y$, then for all $k$ we get $u(\epsilon_k-\epsilon_n)-u(\epsilon_k-\epsilon_j)\neq 0$ for any $j>n$. For $k$'s such that $p_k\neq 0$, then Eq.~\eqref{eq.arbiinco2} is satisfied if and only if $\bra{\epsilon_j}U \ket{\epsilon_k}\bra{\epsilon_k}U^\dagger\ket{\epsilon_n}$ is zero for $j>n$. This happens only if the unitary transformation applied to the projector of energy, i.e., $U \ket{\epsilon_k}\bra{\epsilon_k}U^\dagger$, is a projector of energy such that the off-diagonal elements (with respect to the energy basis) are zero, i.e., $U$ maps $\ket{\epsilon_k}$ to $e^{i\phi_k}\ket{\epsilon_{\pi_k}}$, for $k$ such that $p_k\neq 0$. The complementary subspace generated by $\ket{\epsilon_k}$ with $k$ such that $p_k=0$ does not play any role in the expectation value, thus we can consider an incoherent unitary operator of the form reported in Eq.~\eqref{eq.uinco}.
\end{proof}
From the proof, we note that, in addition to the fact that the optimal incoherent unitary operator still remains defined up to irrelevant phases $\phi_k$, the corresponding optimal permutation can be not unique (it is unique only if $\rho$ has full rank).
We note that a deterministic optimal work extraction can be obtained from a pure state $\rho_{\bar k}= \ket{\epsilon_{\bar{k}}}\bra{\epsilon_{\bar{k}}}$. In this case, by considering a strictly increasing utility function $u(w)$, from Eq.~\eqref{eq.arbiinco} we get $\mathcal U(\rho_{\bar{k}}) = u(\epsilon_{\bar{k}} - \epsilon_1)$. Given a general incoherent state $\rho = \sum_k p_k \ket{\epsilon_k}\bra{\epsilon_k}$, by using the Jensen's inequality, it is easy to see that an agent that is averse to risk ($u(w)$ is concave) prefers to extract the certain amount of work $w_{det}=\mathcal E (\rho_{\bar{k}}) = W(\rho_{\bar{k}},U_E)=\epsilon_{\bar{k}} - \epsilon_1$, i.e.,  $\mathcal U(\rho) < \mathcal U(\rho_{\bar{k}})$, if $W(\rho,U_u)<w_{det}$ or if $\mathcal E(\rho) < w_{det}$. On the other hand, for a risk loving agent ($u(w)$ is convex) $W(\rho,U_u)>w_{det}$ implies that $\mathcal U(\rho) > \mathcal U(\rho_{\bar{k}})$ and thus, in this case, the agent prefers to extract work from the state $\rho$ instead of the deterministic work extraction from $\rho_{\bar{k}}$.

Furthermore, from Eq.~\eqref{eq.arbiinco}, we can give a sufficient condition such that $\mathcal U(\rho)>0$, in analogy with Theorem~\ref{theo2}. For simplicity, we assume that $u(x)$ is a strictly increasing function and $u(0)=0$. If there are $n$ and $k>n$ such that $p_n u(\epsilon_n-\epsilon_k) + p_k u(\epsilon_k-\epsilon_n)>0$, we can consider the permutation $\pi_k$ such that $\pi_k=n$, $\pi_n=k$ and $\pi_s=s$ for $s\neq n,k$. Thus, since $u(0)=0$, for this permutation we get $\langle u(w)\rangle_{\pi_k} = p_n u(\epsilon_n-\epsilon_k) + p_k u(\epsilon_k-\epsilon_n)>0$. By noting that $u(\epsilon_n-\epsilon_k)< u(0)=0$ and $u(\epsilon_k-\epsilon_n)> u(0)=0$, we get:
\begin{theorem}
$\mathcal U(\rho) >0 $ if there are some $n$ and $k>n$ such that
\begin{equation}
\frac{p_k}{p_n} > \frac{\abs{u(\epsilon_n-\epsilon_k)}}{u(\epsilon_k-\epsilon_n)}
\end{equation}
\end{theorem}
By considering that for the exponential utility function $u(w)$ of Eq.~\eqref{eq.uteqxpo} we get $\abs{u(\epsilon_n-\epsilon_k)}/u(\epsilon_k-\epsilon_n)=e^{r(\epsilon_k-\epsilon_n)}$, we note that the result is in perfect agreement with Theorem~\ref{theo2}.

We now proceed by discussing how the optimal value $\mathcal U(\rho)$ differs from the ergotropy $\mathcal E (\rho)$ as done for the exponential utility function. Given an arbitrary utility function $u(w)$, we write it as $u(w) = w  + \delta(w)$, where $\delta (w) = u(w) - w$ depends on some parameter $r$ and we assume that the agent tends to be neutral to risk, i.e., $\delta (w)\to 0$ as $r\to 0$. Then, there is a neighborhood of zero such that if $r\in (\delta_-,\delta_+)$ we get $U_u=U_E$ and thus $\mathcal U(\rho) =  \mathcal E (\rho) + \langle \delta(w) \rangle_E$. By considering $u'(0)>0$, we can redefine $u(w)$ by multiplying it by a positive constant such that we get the Taylor series around zero $\delta(w) = u''(0) w^2/2 + \cdots$. Thus, for  $r\in (\delta_-,\delta_+)$ we have
\begin{equation}\label{eq.usmallrarbi}
\mathcal U (\rho) = \mathcal E(\rho) - \frac{r_A(0)}{2}\langle w^2 \rangle_E + \cdots\,,
\end{equation}
where $r_A(0)$ is the absolute risk aversion in Eq.~\eqref{eq. RA} evaluated at $w=0$. Then, Eq.~\eqref{eq.usmallrarbi} generalizes Eq.~\eqref{eq.usmallr} achieved for an exponential utility function to an arbitrary one. Similarly to the exponential case, given a set of states, the selection of the initial state with largest $\mathcal U$ is determined by the ergotropy and its higher moments through the absolute risk aversion $r_A(0)$ (in first approximation). Finally, to conclude our analysis, let us further discuss the conditions such that $\mathcal U(\rho')>\mathcal U (\rho)$. To do this we focus on the case $d=2$. In general, the utility function reads $u(w) = u_e(w) + u_o(w)$, where $u_e(w)=u_e(-w)$ and $u_o(w)=-u_o(-w)$, defined as $u_{e,o}(w) = (u(w)\pm u(-w))/2$. For simplicity we consider $u(0)=u_e(0)=0$. Given a state $\rho=p \ket{\epsilon_1}\bra{\epsilon_1} + (1-p) \ket{\epsilon_2}\bra{\epsilon_2}$, we get the NOT permutation as the optimal one and
\begin{equation}
\mathcal U(\rho) = u_e(\epsilon_2-\epsilon_1) + (1-2p)u_o(\epsilon_2-\epsilon_1)
\end{equation}
if $u_e(\epsilon_2-\epsilon_1) + C(p)>0$, where $C(p)\equiv(1-2p)u_o(\epsilon_2-\epsilon_1)$, or the identity permutation and $\mathcal U(\rho)=0$ otherwise. Given another state $\rho'=p' \ket{\epsilon_1}\bra{\epsilon_1} + (1-p') \ket{\epsilon_2}\bra{\epsilon_2}$, if $\mathcal U(\rho)>0$ and $\mathcal U(\rho')>0$, we get the general condition
\begin{equation}
\mathcal U (\rho') > \mathcal U(\rho) \Leftrightarrow C(p')>C(p)\,.
\end{equation}
For the exponential utility function, $C(p')>C(p)$ is equivalent to $p>p'$. In particular this general condition is equivalent to the majorization condition in Eq.~\eqref{eq.compare} when this can be applied, e.g., for $u_1+u_2 = u'_1+u'_2$. However, for an arbitrary utility function, $C(p')>C(p)$ is not equivalent to $p>p'$. For instance, we can consider a quadratic utility function $u(w) = a w + b w^2$. In this case, for $a<0$ we get that $C(p')>C(p)$ is equivalent to $p>p'$, and for $a=0$ we get $\mathcal U (\rho') = \mathcal U (\rho)$ for any $\rho$ and $\rho'$ such that $\mathcal U(\rho)>0$ and $\mathcal U(\rho')>0$. This shows how the majorization condition does not hold for an arbitrary utility function, i.e., $\mathcal U(\rho)$ and $\mathcal E_{CE}(\rho)$ are not Schur-concave functionals of $\{u_k\}$. Thus, finding a general condition that holds for an arbitrary utility is extremely difficult if not downright impossible, although some conditions can be found for specific utility functions. In general, depending on the specific utility function, the only way to see if $\mathcal U (\rho') > \mathcal U(\rho)$ can be to explicitly calculate $\mathcal U(\rho)$ and $\mathcal U(\rho')$ from Eq.~\eqref{eq.arbiinco} by finding the corresponding optimal permutations.

\section{Initial quantum coherence}\label{sec.coe}
If there is initial quantum coherence in the energy basis, i.e., $[\rho,H]\neq0$, for the unitary cycle with unitary time-evolution $U$, we get the quasiprobability distribution~\cite{Francica22,Francica222}
\begin{eqnarray}
\nonumber p_q(w,\rho,U)&=&\sum_{k,j,i} \text{Re}\bra{\epsilon_i}\rho\ket{\epsilon_j}\bra{\epsilon_j}U^\dagger \ket{\epsilon_k}\bra{\epsilon_k}U \ket{\epsilon_i}\\
 && \times\delta(w-q\epsilon_i-(1-q)\epsilon_j+\epsilon_k)\,.
\end{eqnarray}
In particular, it reduces to the quasiprobability distribution of Ref.~\cite{Allahverdyan14} for $q=0,1$ and the one of Ref.~\cite{Solinas15} for $q=1/2$.
We aim to calculate the optimal value in Eq.~\eqref{eq.utilitymax}, where the average in Eq.~\eqref{eq.utilitymax} is now calculated with respect to the quasiprobability distribution $p_q(w,\rho,U)$ instead of the probability distribution $p(w,\rho,U)$.
We focus on the exponential utility function in Eq.~\eqref{eq.uteqxpo}. We  get
\begin{equation}
\langle u(w) \rangle = \frac{1}{r}\left(1- \text{Re}\Tr{U v(q H)\rho v((1-q)H) U^\dagger v(-H)}\right)\,.
\end{equation}
Of course, for $r=0$ we get the ergotropy $\mathcal E(\rho)$ and in general we still get $\mathcal U (\rho)\geq 0$, since we get $\langle u(w)\rangle =0 $ for the identity $U=I$.
To find the optimal $U_u$, we consider
\begin{equation}
\delta \langle u(w) \rangle \propto \text{Re}\Tr{X [Uv(q H)\rho v((1-q)H) U^\dagger,v(-H)]}\,.
\end{equation}
The variation $\delta \langle u(w) \rangle$ is zero and $\langle u(w) \rangle$ is maximum if there exists a unitary operator $U=U_u$ which reads
\begin{equation}\label{eq.coe}
U_u = \sum_{k=1}^d e^{i\phi_k} \ket{\epsilon_k}\bra{u_k(q)}\,,
\end{equation}
where we define $\ket{u_k(q)}$ and $u_k$ such that $A_q\ket{u_k(q)}=u_k\ket{u_k(q)}$, where $A_q=v(q H)\rho v((1-q)H)$ and $u_k\geq u_{k+1}$.
In this case, we get the optimal expected utility of the form in Eq.~\eqref{eq.mainres}, with the new $u_k$'s defined above.
We have that $u_k$'s are nonnegative and do not depend on $q$.
\begin{proof}
To prove it, let us show that the eigenvalues $u_k$ of $A_q$ do not depend on $q$. It is enough to note that $A_q=e^{-rqH} \rho e^{-r H} e^{rqH}$ is a similarity transformation of  $A_0=\rho e^{-rH}$, thus $A_q$ has the same eigenvalues of  $A_0$, namely $u_k$'s do not depend on $q$. To prove that they are real, we consider $q=1/2$, in this case $A_{1/2}=A_{1/2}^\dagger$ and they are also nonnegative since $A_{1/2}= B B^\dagger$, where $B=e^{-rH/2}\rho^{1/2}$.
\end{proof}
However, since $A_q$ and $A_0$ are related by a similarity transformation, we get
\begin{equation}
\ket{u_k(q)}\propto e^{-rqH}\ket{u_k(0)}\,,
\end{equation}
then the states $\ket{u_k(q)}$ in general are not mutual orthogonal, $U_u$ in Eq.~\eqref{eq.coe} is not unitary and $A_q\neq \sum u_k \ket{u_k(q)}\bra{u_k(q)}$, except for $q=1/2$.
In general only for $q=1/2$, $A_{1/2}=A_{1/2}^\dagger$ and $\ket{u_k(1/2)}$'s are mutual orthogonal, so that $U_u$ in Eq.~\eqref{eq.coe} is a unitary operator.
Then, the maximum value $\mathcal U (\rho)$ in Eq.~\eqref{eq.mainres} (with $u_k$ sorted eigenvalues of $A_q$) can be achieved with the optimal unitary operator $U_u$ in Eq.~\eqref{eq.coe} only for the quasiprobability representation achieved for $q=1/2$. 
Furthermore, in this case we note that the sufficient conditions in Eqs.~\eqref{eq.compareweak} and~\eqref{eq.compare} still hold with $u_k$  defined above. 
To get some insights for $q\neq 1/2$, we write
\begin{eqnarray}
\langle u(w)\rangle &=& \frac{1}{r}\left(1- \text{Re}\Tr{S_q e^{-\frac{r}{2}H}\rho e^{-\frac{r}{2}H} S_q^{-1} v(-H)}\right)\\
&=& \frac{1}{r}\left(1- \sum_{k,j} u_j x_{jk} e^{r\epsilon_k}\right)\,,
\end{eqnarray}
where we have defined $S_q = U e^{-qrH}e^{\frac{r}{2}H}$, and $x_{jk}=\text{Re}\bra{\epsilon_k}S_q \ket{u_j(1/2)}\bra{u_j(1/2)} S_q^{-1} \ket{\epsilon_k}$. We have that $\sum_j x_{jk}=\sum_k x_{jk}=1$, so that if $x_{jk}\geq 0$ for all $j$ and $k$ then we get that $x_{jk}$'s are the entries of a doubly stochastic matrix. Thus, from the Birkhoff's theorem, the matrix with entries $x_{jk}$ can be expressed as a convex combination of permutation matrices. Then, the expected utility can be expressed as
\begin{equation}
\langle u(w) \rangle = \frac{1}{r}\left(1- \sum_{\alpha} \theta_\alpha \sum_{k} u_{\pi^{(\alpha)}_k} e^{r\epsilon_k}\right)\,,
\end{equation}
where $\theta_\alpha$'s (which depend on the state $\rho$, the unitary operator $U$ and the parameter $q$) sum to one.
The optimal value reads
\begin{equation}\label{eq.sup2}
\mathcal U(\rho)= \frac{1}{r}\left(1- \sum_{\alpha} \bar{\theta}_\alpha \sum_{k} u_{\pi^{(\alpha)}_k} e^{r\epsilon_k}\right)\,,
\end{equation}
which depends on $q$ only through the coefficients $\bar{\theta}_\alpha$.
However, in a neighborhood of $q=1/2$, we have that $x_{kk}\geq 1$ and $x_{j k}\leq 0$ for all $k$ and $j\neq k$, and all the equalities hold only for $q=1/2$.
For instance, let us focus on the case $d=2$, where we have $x_{12}=x_{21}$ and $x_{11}=x_{22}$. For $q\neq 1/2$, we find $x_{12}<0$ and so Eq.~\eqref{eq.sup2} is an affine combination with $\bar\theta_I + \bar{\theta}_{NOT}=1$, where $\bar{\theta}_{NOT}<0$ and $\bar\theta_I>1$.
\begin{proof}
To prove it, we note that for any $d$, for $U=U_u$, we get $\partial_q x_{jk}|_{q=1/2}=0$ for all $j$ and $k$. Thus, $q=1/2$ is a stationary point of all the entries $x_{jk}$. In particular, for $q=1/2$, $x_{kk}=1$ and $x_{jk}=0$ for all $k$ and $j\neq k$. Due to the symmetry under the transformation $q\mapsto 1-q$, the stationary point can be a minimum or a maximum. We find that $\partial^2_q x_{jk}|_{q=1/2}=2 r^2 (\bra{u_k(1/2)} H^2\ket{u_k(1/2)}\delta_{k,j}-\abs{\bra{u_k(1/2)} H\ket{u_j(1/2)}}^2)+\cdots$. Actually, the omitted part is equal to $-2 (\bra{\epsilon_k(1/2)} M_{1/2}^2\ket{\epsilon_k(1/2)}\delta_{k,j}-\abs{\bra{\epsilon_k(1/2)} M_{1/2}\ket{\epsilon_j(1/2)}}^2)$, where $\partial_q U_u = i M_{q} U_u$, with $M_q^\dagger = M_q$, in detail $M_q= i U_u \partial_q U_u^\dagger$. This contribution has the same form of the one coming from the Hamiltonian, but with opposite sign. However, due to the symmetry under the transformation $q\mapsto 1-q$, we get $\partial_q U_u|_{q=1/2}=0$, i.e., the omitted part is equal to zero. Then $\partial^2_q x_{kk}|_{q=1/2}>0$ (since we get a variance, which is positive) and $\partial^2_q x_{jk}|_{q=1/2}<0$ for $k\neq j$, which completes the proof. This means that, e.g., for $d=2$, $\bar{\theta}_I$ gets its minimum value for $q=1/2$ that is equal to one and $\bar{\theta}_{NOT}$ gets its maximum value for $q=1/2$ that is equal to zero. Then, $\bar\theta_I>1$ and $\bar\theta_{NOT}<0$ for $q\neq 1/2$. 
\end{proof}
Then, in this case we find that the decomposition in Eq.~\eqref{eq.sup2} holds for any $d$, with $\sum_\alpha \bar{\theta}_\alpha=1$, where $\bar\theta_I\geq 1$ and $\bar\theta_\alpha\leq 0$ for $\alpha\neq I$, and all the equalities hold only for $q=1/2$ (the proof in Appendix~\ref{app.deco}).
This affine decomposition implies that the optimal expected utility $\mathcal U(\rho)$ in the function of $q$ gets its minimum value for $q=1/2$. This means that an agent, who selects a quasiprobability representation, obtains at least the optimal value $\mathcal U(\rho)$ achieved for $q=1/2$.
\subsection{Coherent contribution}
We define the coherent contribution to the expected utility as
\begin{equation}
\mathcal U_c(\rho)= \mathcal U(\rho)-\mathcal U(\Delta(\rho))\,.
\end{equation}
For $q=1/2$, we have that $\mathcal U_c(\rho)\geq0$, thus the agent prefers the state with quantum coherence instead of the dephased $\Delta(\rho)$. This generalizes the $r=0$ result $\mathcal E_c(\rho)= \mathcal E(\rho)-\mathcal E(\Delta(\rho))\geq 0$ in Ref.~\cite{Francica20}.
\begin{proof}
To prove it, we use Eq.~\eqref{eq.compare}. We note that $\Delta(\rho)$ is a unital map, i.e., $\Delta(I)=I$, and
\begin{equation}
\Delta(v(q H)\rho v((1-q)H))=\Delta(\rho)e^{-r H}\,,
\end{equation}
then $u_k$ majorizes $u'_k$, where $u'_k$ are the sorted eigenvalues of $\Delta(\rho)e^{-r H}$. For $r\neq 0$, the condition  $\sum_{j=1}^d (u_j-u_j')=0$ is equivalent to $\Tr{\rho v(H)}=\Tr{\Delta(\rho) v(H)}$ which is satisfied. Thus by using Eq.~\eqref{eq.compare}, we get $\mathcal U_c(\rho)\geq 0$.
\end{proof}
A similar definition can be given for the coherent contribution to the certainty equivalent
\begin{equation}
\mathcal E_{CE,c}(\rho)= \mathcal E_{CE}(\rho)-\mathcal E_{CE}(\Delta(\rho))\,,
\end{equation}
such that $\mathcal E_{CE,c}(\rho)\geq 0$ for $q=1/2$.
For $r=0$, we get the ergotropy and so $\mathcal U_c(\rho)=\mathcal U(U_I\rho U_I)$, where $U_I$ is the optimal incoherent unitary operator for the initial state $\Delta(\rho)$. However, for $r\neq 0$ a similar equality does not hold since, while the work is additive under composition of unitary cycles, i.e.,  if $U=U_c U_I$ we get $W (\rho, U)= W(U_I \rho U_I,U_c) + W(\rho,U_I)$, the expected utility $\langle u(w)\rangle$ (or the certainty equivalent) is not (when $u(w)$ is nonlinear).
Finally, to study the behavior of the utility as a function of initial quantum coherence, we focus on a qubit in the state
\begin{equation}
\rho= p \ket{\epsilon_1}\bra{\epsilon_1}+ (1-p) \ket{\epsilon_2}\bra{\epsilon_2} + c \ket{\epsilon_1}\bra{\epsilon_2}+c^*\ket{\epsilon_2}\bra{\epsilon_1}\,,
\end{equation}
where $|c|\leq \sqrt{p(1-p)}$, and $H=\epsilon\ket{\epsilon_2}\bra{\epsilon_2}$. The case $c=0$ was discussed in Sec.~\ref{sec.exputid2}. For $q=1/2$ the coherent contribution can be exactly calculated and reads
\begin{equation}
\mathcal U_c(\rho)= \left(|\eta|-\sqrt{4|c|^2e^{r\epsilon}+\eta^2}\right)\frac{e^{-r \epsilon}-1}{2r}\,,
\end{equation}
where $\eta=p(1+e^{r\epsilon})-1$. Thus, we deduce that, in this particular case, $\mathcal U_c(\rho)$ increases as the initial quantum coherence $|c|$ increases for any $r$.
In the end, we also note that the results of an ensemble of $n$ batteries of Sec.~\ref{sec.ensemble} can be easily generalized in the presence of quantum coherence for $q=1/2$ (see Appendix~\ref{app.battecohe}).

\section{Conclusion}
An agent that is non-neutral to risk performs the selection of a work extraction procedure with a given work statistics by looking on the expected utility, which characterizes the satisfaction of the selection. Here, we answer to the question: given a Hamiltonian, from which state does a risk non-neutral agent prefer to extract work? To answer to this question, we introduce the optimal expected utility by maximizing over all unitary cycles, generalizing the concept of ergotropy. The selection is made thanks to the expected utility hypothesis, comparing the optimal expected utility among the available initial states and choosing the state with the highest one. For incoherent initial states, we find the form of the optimal expected utility, which is achieved by an optimal incoherent unitary transformation. In the case of an exponential utility function, we find that majorization gives a sufficient condition to perform the selection. The optimal incoherent unitary cycle is investigated with the help of some examples, which are a qubit, a qutrit and two qubits.
Furthermore, we examine an ensemble of quantum batteries, also showing that it is possible to get the maximum certainty equivalent and the ergotropy at the same time for certain unitary cycles. In particular, a general formula for the generating function of the ergotropy cumulants trivially follows from our results.
Interestingly, concerning the effects of the initial quantum coherence, by adding quantum coherence in an initial incoherent state the optimal expected utility increases, so that the agent prefers the state with initial quantum coherence instead of the incoherent one. For instance, for a qubit the agent prefers the state with more initial quantum coherence, if the populations remain fixed.
In the end, we think that expected utility gives an important criterion on how to optimize work extraction from quantum batteries in situations where the fluctuations are relevant. For instance, a single realization of the unitary cycle, which gives the maximum average work (i.e., the ergotropy), can give a random work far from the average value. In this case, searching an alternative unitary cycle that optimizes the extraction process by also looking at reducing fluctuations can be vital for agents averse to risk, and one way to achieve this is to maximize the expected value of a suitable utility rather than average work.

\subsection*{Acknowledgements}
The authors acknowledge financial support from the project BIRD 2021 "Correlations, dynamics and topology in long-range quantum systems" of the Department of Physics and Astronomy, University of Padova, from the European Union - Next Generation EU within the National Center for HPC, Big Data and Quantum Computing (Project No. CN00000013, CN1 - Spoke 10 Quantum Computing) and  from the Project "Frontiere Quantistiche" (Dipartimenti di Eccellenza) of the Italian Ministry for Universities and Research.

\appendix

\section{Maximum work extractable from an ensemble of quantum batteries}\label{app.AlickiFannes}
Let us focus on the average work and derive the result of Eq.~\eqref{eq.Alicki-Fannes}.
The average work can be obtained from Eq.~\eqref{eq.exp-rw} by considering that $\langle e^{-rw}\rangle\sim 1-r\langle w\rangle$ as $r\to 0$, getting
\begin{eqnarray}\label{eq.avework}
\nonumber\langle w\rangle &=& \sum_{n_1\ldots n_d|\sum_i n_i=n} \sum_{\bar n_1 \ldots \bar n_d |\sum_i \bar n_i = n} p_{n_1\ldots n_d}p_{\bar n_1\ldots \bar n_d|n_1 \ldots n_d}\\
&&\times  \sum_i(n_i-\bar n_i)\epsilon_i\,.
\end{eqnarray}
In the limit $n\to \infty$, the sums over $\{n_i\}$ give integrals over $\{\tilde p_i\}$. Thus, in the first sum of Eq.~\eqref{eq.avework}, only a distribution $\tilde p= \tilde p^*=\{\tilde p^*_i\}$ gives the dominant contribution, which makes stationary the Lagrangian $\tilde F[\tilde p]=D(\tilde p || p) + D(\bar p^*) + \mu \sum_i \tilde p_i$, where we have introduced the Lagrange multiplier $\mu$ such that $\sum_i \tilde p_i = 1$ and $\bar p^*=\{\bar p^*_i\}$ is a stationary point which gives the dominant contribution of the second sum of Eq.~\eqref{eq.avework}. Before to give the precise definition of $\bar p^*$, we note that the maximization of the asymptotic value of the average work in Eq.~\eqref{eq.avework} gives a condition that fixes the conditional probabilities $p_{\bar n_1\ldots \bar n_d|n_1 \ldots n_d}$, i.e., the operator $U_{\{n_i\}}$, only for the occupation numbers $\{n_i = n \tilde p^*_i\}$.
Thus, we focus on $\{n_i = n \tilde p^*_i\}$. We start to note that there are unitary operators $U$ giving
\begin{equation}\label{eq.satuprob}
p_{\bar n_1\ldots \bar n_d|n_1 \ldots n_d}=\prod_i \delta_{\bar n_i,n'_i}\,,
\end{equation}
where the set $\{n'_i\}$ is defined such that the number $W_{\bar n_1 \ldots \bar n_d | n_1 \ldots n_d}$ is equal to $W_{n_1\ldots n_d}$ for $\{\bar{n}_i\}=\{n'_i\}$ and zero otherwise. Thus, as $n\to \infty$,  $n'_i= n p'_i$ where $p'=\{p'_i\}$ has the same Shannon entropy of $\tilde p^*$, $H(p')=H(\tilde p^*)$. We aim to show that the optimal $U_{\{n_i\}}$ gives conditional probabilities of the form of Eq.~\eqref{eq.satuprob}.
By considering only the term $\tilde p = \tilde p^*$ in the first sum in Eq.~\eqref{eq.avework}, and noting that $p_{\bar n_1\ldots \bar n_d|n_1 \ldots n_d}\sim e^{-n D(\bar p)}$, we get
\begin{equation}\label{eq.prov}
 \sum_{\bar n_1 \ldots \bar n_d |\sum_i \bar n_i = n}   e^{-n D(\bar p)} \sum_i\bar n_i\epsilon_i   \sim   n e^{-n D(\bar p^*)} \sum_i \bar{p}^*_i\epsilon_i\,,
\end{equation}
where $\bar p^*$ makes stationary the Lagrangian $\bar F [\bar p] = D(\bar p) + \bar \mu \sum_i \bar p_i $, where we have introduced the Lagrange multiplier $\bar\mu$ such that $\sum_i \bar p_i = 1$. We note that the left side of Eq.~\eqref{eq.prov} is non-smaller than $\sum_{i=1}^{W_{\{n_i\}}} E_i / W_{\{n_i\}}$, where $E_i$ are the energies of the states $\ket{\{i_k\}}$ sorted such that $E_i\leq E_{i+1}$. By considering $\epsilon_i\geq 0$, we get that Eq.~\eqref{eq.prov} is larger than $\epsilon_{2}$ and thus different from zero as $n\to \infty$. This implies that the stationary point $\bar p^*$ is such that $D(\bar p^*)=0$, otherwise Eq.~\eqref{eq.prov} exponentially decays to zero as $n\to \infty$. By noting that $W_{\bar n_1 \ldots \bar n_d| n_1 \ldots n_d} \sim e^{n(H(\tilde p^*)-D(\bar p))}$, since $D(\bar p^*)=0$, for $\{\bar n_i = n \bar p^*_i\}$ we get $W_{\bar n_1 \ldots \bar n_d| n_1 \ldots n_d} \sim e^{nH(\tilde p^*)}$, thus $W_{\bar n_1 \ldots \bar n_d| n_1 \ldots n_d} \sim W_{ n_1 \ldots n_d}$ with $\{n_i = n \tilde p^*_i\}$. Then, we deduce that $\bar p^* = p'$, since $p'$ is a general distribution probability defined such that $W_{n'_1 \ldots n'_d| n_1 \ldots n_d} \sim W_{ n_1 \ldots n_d}$.
Thus, $p_{\bar n_1\ldots \bar n_d|n_1 \ldots n_d}$ decays exponentially as $\bar p_i$ is different from $p'_i$, and we get
\begin{equation}\label{eq.001}
 \sum_{\bar n_1 \ldots \bar n_d |\sum_i \bar n_i = n} p_{\bar n_1\ldots \bar n_d|n_1 \ldots n_d} \sum_i\bar n_i\epsilon_i  \sim n \sum_i p'_i\epsilon_i\,.
\end{equation}
Furthermore, since $D(\bar p^*)=0$, from $\delta \tilde F[\tilde p] = 0$ we get the stationary point $\tilde p^*=p$, and so $U_{\{n_i\}}: \mathcal H_{\{n_i\}}\to \mathcal H_{\{n'_i\}}$ for the typical values of the occupation numbers $\{n_i=np_i\}$.
Thus, by considering only the stationary points in the sums in Eq.~\eqref{eq.avework}, the average extracted work can be expressed as
\begin{equation}\label{eq.aveworkasym}
\langle w\rangle \sim n \sum_i(p_i-p'_i)\epsilon_i\,.
\end{equation}
The value of $p'={p'}^*$ that maximizes the asymptotic formula in Eq.~\eqref{eq.aveworkasym} will make stationary the Lagrangian $F'[p'] = \sum_i p'_i \epsilon_i + \lambda H(p') + \nu \sum_i p'_i$, where we have introduced the Lagrange multipliers $\lambda$ such that $H(p')=H(p)$ and $\nu$ such that $\sum_i  p'_i = 1$. By requiring that $\delta F'[p']=0$, we get
\begin{equation}
{p_i'}^* = \frac{e^{-\beta \epsilon_i}}{Z}
\end{equation}
where $Z=\sum_i e^{-\beta \epsilon_i}$ with $\beta$ solution of $H({p'}^*)=H(p)$. This means that the final average energy is asymptotically equal to the one of the completely passive state, and so we get the result in Eq.~\eqref{eq.Alicki-Fannes}.

\section{Expressions for small $r$ }\label{app.smallr}
For small $r$, by looking for a solution $\tilde p^*_i = \sum_{s=0} \tilde p_{s,i} r^s$ of Eq.~\eqref{eq.tildep02}, and $\mu = \sum_{s=0} \mu_s r^s$, we get $p_{0,i}=p_i$, $\mu_0=-1$, thus as $r\to 0$ we get the conventional result for the optimal ${p'}^*$ with $\beta$ solution of $H({p'}^*)=H(p)$. At the first order we get
\begin{eqnarray}
\tilde p_{1,i} &=& \sum_j \frac{\partial \beta}{\partial \tilde p_i}\frac{\partial}{\partial \beta}\left(\frac{e^{-\beta\epsilon_j}}{Z}\right)\bigg|_{\tilde p_i=p_i}\epsilon_j p_i-\epsilon_ip_i-\mu_1p_i \,,\\
\mu_1 &=& \sum_{i,j} \frac{\partial \beta}{\partial \tilde p_i}\frac{\partial}{\partial \beta}\left(\frac{e^{-\beta\epsilon_j}}{Z}\right)\bigg|_{\tilde p_i=p_i}\epsilon_j p_i-\sum_i\epsilon_ip_i\,,
\end{eqnarray}
from which
\begin{equation}
{p'_i}^* = \left(\frac{e^{-\beta\epsilon_i}}{Z}\right)\bigg|_{\tilde p_i=p_i}+\sum_j \frac{\partial \beta}{\partial \tilde p_j}\frac{\partial}{\partial \beta}\left(\frac{e^{-\beta\epsilon_i}}{Z}\right)\bigg|_{\tilde p_j=p_j} \tilde p_{1,j}r+ O(r^2)
\end{equation}
and
\begin{equation}
f[\tilde p^*] = \sum_i \frac{\tilde p_{1,i}^2}{2p_i}r^2 +r\sum_i(p_i+\tilde p_{1,i} r-{p'_i}^*)\epsilon_i+ O(r^3)\,.
\end{equation}

\section{Affine decomposition}\label{app.deco}
We consider the $d\times d$ matrix $x$ with entries $x_{jk}$, such that $\sum_{j}x_{jk}=\sum_{k}x_{jk}=1$, $x_{kk}\geq 1$ and $x_{j k}\leq 0$ for all $k$ and $j\neq k$. We aim to show that there is an affine combination of permutation matrices $P^{(\alpha)}$ such that
\begin{equation}\label{eq.app.ansatz}
x = \sum_\alpha \theta_\alpha P^{(\alpha)}\,,
\end{equation}
with $\sum_\alpha \theta_\alpha = 1$, $\theta_I \geq 1$ and $\theta_{\alpha\neq I}\leq 0$, with $P^{(I)}=I$. For $d=2$, Eq.~\eqref{eq.app.ansatz} is true since we have
\begin{equation}
x = \left(
      \begin{array}{cc}
        \theta_I & \theta_{NOT} \\
        \theta_{NOT}& \theta_I \\
      \end{array}
    \right) = \theta_I P^{(I)} + \theta_{NOT} P^{(NOT)} \,.
\end{equation}
For $d=3$, we get $3!=6$ permutations, which are $I:(1,2,3)$, $2:(1,3,2)$, $3:(2,1,3)$, $4:(2,3,1)$, $5:(3,1,2)$ and $6:(3,2,1)$. We get that Eq.~\eqref{eq.app.ansatz} is true by choosing
\begin{eqnarray}
\label{eq.app.1}&&\theta_2 = x_{11}-\theta_I\,, \quad \theta_3 = x_{33} - \theta_I\,,\quad \theta_{6} = x_{22}-\theta_I\,,\\
\label{eq.app.2}&&\theta_4 = x_{12}-x_{33}+\theta_I\,, \quad \theta_5 = x_{13}-x_{22}+\theta_I
\end{eqnarray}
and $\theta_I$ any real such that $\theta_I \geq \max_k x_{kk}$. By considering $x_{11}\geq x_{22} \geq x_{33}$, we get that $\theta_\alpha$ in Eq.~\eqref{eq.app.1} are trivially nonpositive, and $\theta_\alpha$ in Eq.~\eqref{eq.app.2} are also nonpositive since $\theta_4 \leq x_{23}\leq 0$ and $\theta_5\leq x_{32}\leq 0$. To prove it for any $d$, we define $\theta_I = \max_k x_{kk}= x_{\bar{k}\bar{k}}\geq 1$ and the matrix $\tilde x = (x- \theta_I P^{(I)})/(1-\theta_I)$. We get that the entries are nonnegative $\tilde x_{jk} \geq 0$ and $\sum_j \tilde x_{jk} = \sum_k \tilde x_{jk}=1$, then $\tilde x$ is a doubly stochastic matrix. Since $\tilde x_{\bar{k}\bar{k}}=0$, from the Birkhoff's theorem we get the convex combination
\begin{equation}
\tilde x = \sum_{\alpha\neq I} \tilde\theta_\alpha P^{(\alpha)}\,,
\end{equation}
from which
\begin{equation}
x = \theta_I P^{(I)} + (1-\theta_I) \sum_{\alpha\neq I} \tilde\theta_\alpha P^{(\alpha)}\,,
\end{equation}
which is equal to Eq.~\eqref{eq.app.ansatz} by noting that $\theta_\alpha = (1-\theta_I) \tilde\theta_\alpha \leq 0$ for $\alpha\neq I$.

\section{Ensemble of quantum batteries with initial quantum coherence}\label{app.battecohe}
We focus on $q=1/2$, so that we get
\begin{equation}
\langle e^{-r w}\rangle =  \Tr{\left(e^{-\frac{rH}{2}}\rho e^{-\frac{rH}{2}}\right)^{\otimes n} U^\dagger \left(e^{r H}\right)^{\otimes n} U}\,.
\end{equation}
We can consider the spectral decomposition $e^{-\frac{rH}{2}}\rho e^{-\frac{rH}{2}}=\sum_{i=1}^d u_i \ket{u_i}\bra{u_i}$, where we recall that $u_i\geq 0$. The optimal unitary $U$ maps a state $\otimes_k \ket{u_{i_k}}$ to a state $ \ket{\{\bar{i}_k\}}$. For simplicity, we assume that $u_i$'s are nondegenerate.
Then, by proceeding as for the incoherent case in Sec~\ref{sec.ensemble}, we get the optimal certainty equivalent in Eq.~\eqref{eq.ECEnbatt} with
\begin{equation}
f[\tilde p] = -H(\tilde p) -\sum_i \tilde p_i \ln u_i - r \sum_i {p'_i}^* \epsilon_i\,,
\end{equation}
where  $\tilde p$ is solution of
\begin{equation}\label{eq.tildep02coe}
\ln \tilde p_i - \ln u_i + 1 - r \sum_{j} \epsilon_j  \frac{\partial  {p'_j}^*}{\partial \tilde p_i} + \mu=0\,,
\end{equation}
and ${p'}^*$ is given by Eq.~\eqref{eq.pprime}.



\begin{thebibliography}{99}

\bibitem{campaioli23} F. Campaioli, S. Gherardini, J. Q. Quach, M. Polini, and G. M. Andolina, arXiv:2308.02277 (2023).



\bibitem{Allahverdyan04} A. E. Allahverdyan, R. Balian, and T. M. Nieuwenhuizen, Europhys. Lett. 67, 565 (2004).

\bibitem{vonNeumann} J. von Neumann, and O. Morgenstern, Theory of Games and Economic Behavior (60th Anniversary Commemorative Edition) (Princeton University Press, 2007).


\bibitem{Ducuara23} A. F. Ducuara, P. Skrzypczyk, F. Buscemi, P. Sidajaya, and V. Scarani, Phys. Rev. Lett. 131, 197103 (2023).
\bibitem{Francica23} G. Francica, and L. Dell'Anna, Phys. Rev. E 109, 014112 (2024).

\bibitem{Perarnau-Llobet17} M. Perarnau-Llobet, E. B\"{a}umer, K. V. Hovhannisyan, M. Huber, and A. Acin, Phys. Rev. Lett. 118, 070601 (2017).
\bibitem{Lostaglio18} M. Lostaglio, Phys. Rev. Lett. 120, 040602 (2018).


\bibitem{Francica22} G. Francica, Phys. Rev. E 105, 014101 (2022).
\bibitem{Francica222} G. Francica, Phys. Rev. E 106, 054129 (2022).

\bibitem{Talkner07} P. Talkner, E. Lutz, and P. H\"{a}nggi, Phys. Rev. E 75, 050102(R) (2007).

\bibitem{campisi11} M. Campisi, P. H\"{a}nggi, and P. Talkner, Rev. Mod. Phys. 83, 771 (2011).



\bibitem{bookmicroeco} D. Mas-Colell, M. Winston, and J. Green, Microeconomic Theory, Oxford, Oxford University Press, 1995.
\bibitem{bookmicroeco2} D. Kreps, A Course in Microeconomic Theory, New Jersey, Princeton University Press, 1990.


\bibitem{Morales23} P. A. Morales, J. Korbel, and F. E. Rosas, New J. Phys. 25 073011 (2023).

\bibitem{Allahverdyan2005} A. E. Allahverdyan, and Th. M. Nieuwenhuizen, Phys. Rev. E 71, 046107 (2005).

\bibitem{Francica22ergo} G. Francica, Phys. Rev. E 105, L052101 (2022).

\bibitem{Alicki13} R. Alicki, and M. Fannes, Phys. Rev. E 87, 042123 (2013).



\bibitem{Allahverdyan14} A. E. Allahverdyan, Phys. Rev. E 90, 032137 (2014).
\bibitem{Solinas15} P. Solinas, and S. Gasparinetti, Phys. Rev. E 92, 042150 (2015).

\bibitem{Francica20} G. Francica, F. C. Binder, G. Guarnieri, M. T. Mitchison, J. Goold, and F. Plastina, Phys. Rev. Lett. 125, 180603 (2020).



\end{thebibliography}
\end{document}